\documentclass[12pt]{iopart}
\usepackage{amssymb}

\usepackage{epsfig}
\usepackage{dcolumn}
\usepackage{bm}
\usepackage{hyperref}
\usepackage{latexsym}

\begin{document}

\title[Probability currents as principal characteristics ...]%
{Probability currents as principal characteristics in the statistical
mechanics of nonequilibrium steady states}
\author{R.~K.~P.~Zia and B.~Schmittmann}
\address{Center for Stochastic Processes in Science and Engineering, 
Department of Physics, Virginia Tech, Blacksburg, VA 24061-0435, USA}
\date{\today }

\begin{abstract}
One of the key features of nonequilibrium steady states (NESS) is the
presence of nontrivial probability currents. We propose a general
classification of NESS in which these currents play a central distinguishing
role. As a corollary, we specify the transformations of the dynamic
transition rates which leave a given NESS invariant. The formalism is most
transparent within a continuous time master equation framework since it
allows for a general graph-theoretical representation of the NESS. We
discuss the consequences of these transformations for entropy production,
present several simple examples, and explore some generalizations, to discrete time
and continuous variables.
\end{abstract}

\vspace{-0.8cm}
\ead{\mailto{rkpzia@vt.edu}, \mailto{schmittm@vt.edu}}

\section{Introduction.}
\label{intro}

The characterization and theoretical understanding of non-equilibrium
phenomena forms one of the greatest challenges of current statistical
physics \cite{SZrev,DMrev,INI}. In contrast to systems in thermal
equilibrium, systems far from equilibrium carry nontrivial fluxes of
physical quantities such as particles or energy. 
These fluxes are induced and maintained by
coupling the system to multiple reservoirs, acting as sources and sinks (of
particles or energy, say) for the system. Since systems of this type abound
in biology, chemistry, and engineering, any progress in this area is likely
to have significant impact, well beyond condensed matter physics.

Over the past years, the remarkable richness of non-equilibrium physics has
been illustrated beautifully, through the study of simple models. Yet, a
satisfactory and comprehensive theoretical framework is still missing, even
for the arguably simplest generalizations of equilibrium, namely,
non-equilibrium stationary states (NESS). At the source of the difficulties
lie several features: First, the evolution of many interesting statistical
systems is described by a set of transition rates without the basis of a
known, underlying Hamiltonian. Familiar examples span a wide range, e.g.,
chemical reactions, cell functions, population dynamics, social networks,
and financial markets. Second, when (and if) such system reaches a
stationary state, it will be a NESS in general. For such states, there is no 
\emph{a priori }knowledge of, or a good working hypothesis for, the
configurational probability distribution. By contrast, for systems known to
reach thermal equilibrium, the fundamental hypothesis is highly successful
in providing the stationary distribution, regardless of the details of the
dynamical evolution. Finally, seemingly minor modifications of the
evolutionary rules (i.e., the system dynamics) often lead to dramatic
changes in macroscopic properties, suggesting that \emph{very different}
NESS's can be associated with \emph{just slightly different} dynamics.

To appreciate how profoundly NESS differ from equilibrium systems in this
regard, let us briefly recall the Gibbs-Boltzmann framework for systems in
thermal equilibrium (with a single reservoir). For simplicity, we use the
language of the canonical ensemble here \cite{canon}.
The required inputs are: a set of configurations $%
\mathcal{C}_{1}$, $\mathcal{C}_{2}$, ... of the system (also known as
microstates) and an expression for the internal energy $\mathcal{H}(\mathcal{%
C})$ associated with each configuration $\mathcal{C}$. Then, the statistical
weight, $P^{eq}(\mathcal{C)}$, for a system in thermal equilibrium with a
reservoir is well established, in terms of $\mathcal{H}(\mathcal{C})$ and a
simple parameter - temperature $T$ - associated with the reservoir. All
macroscopic observables now follow as configurational averages, $%
\left\langle A\right\rangle =$ $\sum_{\mathcal{C}}$ $A(\mathcal{C)}P^{eq}(%
\mathcal{C)}$. In particular, a system coupled to a heat bath at inverse
temperature $\beta =1/k_{B}T$ is governed by the familiar Boltzmann
distribution, $P^{eq}(\mathcal{C)}=Z^{-1}\exp [-\beta \mathcal{H}(\mathcal{C}%
)]$.

Of course, all real systems, whether in equilibrium or not, are
fundamentally dynamic in nature, continuously undergoing transitions from
one configuration to another. In that context, stationary distributions,
including $P^{eq}(\mathcal{C)}$, emerge as the long-time limit of a
time-dependent distribution, $P(\mathcal{C};t\mathcal{)}$. For its time
evolution, we will assume that it obeys a master equation, with a given set
of transition rates between the configurations. From a modelling
perspective, it is therefore natural to ask: What choices of dynamic
transition rates will lead to a desired stationary distribution? and: What
modifications of these rates will leave this distribution invariant? For
equilibrium systems, the answer is well known and can be traced to the
property of \emph{detailed balance}, which is related to \emph{microscopic
reversibility}. Specifically, if the rates satisfy detailed balance, the 
\emph{net} probability current between \emph{any pair} of configurations
vanishes in the steady state. As a result, $P^{eq}(\mathcal{C)\equiv }%
\lim_{t\rightarrow \infty }P(\mathcal{C};t\mathcal{)}$ can be expressed in
terms of \emph{ratios} of these rates, and the long-time limit remains
invariant under any modification of the dynamics which preserves these 
ratios. Indeed, Monte Carlo simulation studies of equilibrium systems rely
heavily on this property. Since the full configurational sum $\sum_{\mathcal{%
C}}$ is computationally inaccessible, an importance sampling of
configuration space is achieved dynamically, by constraining the transition
rates such that $P(\mathcal{C};t\mathcal{)}$ eventually approaches the
desired equilibrium distribution.

In this article, we begin by addressing two fundamental questions. First, is
there a general procedure to find the stationary solution(s) of a master
equation even if the transition rates violate detailed balance? Second, can
we specify the class of transformations of the rates which leave this
stationary distribution invariant? The answer to the first question is not
new, expressing the solution in terms of directed trees. Since it does not
seem to be widely known, we will provide a brief review here, restricting
ourselves to systems with (\emph{i}) a finite (but arbitrary) number of
configurations, (\emph{ii}) time-independent transition rates, and (\emph{iii%
}) ergodicity. In this framework, we discuss the importance of irreversible
loops (associated with the transition rates) in configuration space, the
role they play in the non-trivial probability currents, and how equilibrium
distributions are recovered in case all loops are reversible. To answer the
second question, we proposed \cite{ZS-JPAL} a general classification of NESS
in terms of the stationary configurational probabilities $P^{\ast }(\mathcal{%
C})\;$and the (stationary)\ probability currents connecting them, $K^{\ast }(%
\mathcal{C}^{\prime },\mathcal{C})$ - denoted in short by $\{P^{\ast
},K^{\ast }\}$. A given set $\{P^{\ast },K^{\ast }\}$ defines a particular
NESS, and allows us to compute all physical currents - mass, energy, etc -
for this state. We then discuss the set of transformations of the transition
rates which leave $\{P^{\ast },K^{\ast }\}$ invariant. In other words, we
describe how to generalize the ``detailed balance condition'' to NESS.

In the reminder of this article, we explore some consequences of our
postulate and present several specific examples, to illustrate the very
general and formal framework proposed. In a concluding section, we provide a
summary and outlook, including some remarks on generalizations to discrete
time and continuous configuration space. The Appendices are devoted to
technical details. 

\section{The master equation and its associated steady state }
\label{master}

We first establish the mathematical framework for our analysis. Consider a
general continuous-time dynamics, with a discrete, finite configuration
space. We assume that every configuration can be reached from every other
one, so that the system is ergodic. Labelling the configurations in some
arbitrary manner as $\mathcal{C}_{1}$, $\mathcal{C}_{2}$, .., $\mathcal{C}%
_{N}$, we are interested in $P\left( \mathcal{C}_{i};t\right) $, the
probability to find the system in configuration $\mathcal{C}_{i}$ at time $t$%
. Its evolution is governed by a set of transition rates $w\left( \mathcal{C}%
_{j}\rightarrow \mathcal{C}_{i}\right) $, for the system to change from
configuration $\mathcal{C}_{j}$ to configuration $\mathcal{C}_{i}$, per unit
time. To simplify notation, let $P_{i}(t)$ stand for $P\left( \mathcal{C}%
_{i};t\right) $ and $w\left( \mathcal{C}_{j}\rightarrow \mathcal{C}%
_{i}\right) $ be denoted 
by $w_{i}^{j}$ \cite{Einstein}. All $w_{i}^{j}$ are real, non-negative,
and assumed to be time-independent. In general, ``forward''\ and
``backward'' rates differ, i.e., $w_{i}^{j}\neq w_{j}^{i}$. If one of them
vanishes while the other remains nonzero, we will call the corresponding
transition uni-directional. In terms of these $w$'s, the master equation
simply expresses the rate of change of $P_{i}(t)$ as the system makes
transitions in and out of $\mathcal{C}_{i}$: 
\begin{equation}
\partial _{t}P_{i}(t)=\sum_{j\neq i}\left[
w_{i}^{j}P_{j}(t)-w_{j}^{i}P_{i}(t)\right]   \label{me}
\end{equation}%
This is often more conveniently written in matrix form 
\begin{equation}
\partial _{t}P_{i}(t)=\sum_{j=1}^{N}W_{i}^{j}P_{j}(t)  \label{me-W}
\end{equation}%
where we have introduced the $N\times N$ matrix $W$ via 
\begin{equation}
W_{i}^{j}=\left\{ 
\begin{array}{c}
w_{i}^{j} \\ 
-\sum_{k\neq j}w_{k}^{j}%
\end{array}%
\right. \ \rm{if} \quad
\begin{array}{c}
i\neq j \\ 
i=j%
\end{array}%
\label{def-W}
\end{equation}%
We note that $W_{i}^{j}$ is a stochastic matrix (in the continuous-time
formalism) since it satisfies (i) $W_{i}^{j}\geq 0$ for all $i\neq j$, and
(ii)$\;\sum_{i}W_{i}^{j}=0$ for all $j$. The second condition ensures that,
once normalized, $P_{i}(t)$ remains so for all subsequent times.

The right hand side of Equation (\ref{me}) can be expressed in terms of the (net) 
\emph{probability} currents $K_i^j$, from configuration $\mathcal{C}_j$ into
configuration $\mathcal{C}_i$, 
\begin{equation}
K_i^j\left( t\right) \equiv w_i^jP_j(t)-w_j^iP_i(t)  \label{P-current}
\end{equation}
so that the master equation simply states the conservation of probability: $%
\partial _tP_i(t)=\sum_{j\neq i}K_i^j\left( t\right) $.

Since the dynamics is ergodic, Equation (\ref{me}) has a unique stationary
solution, $P_i^{*}\equiv \lim_{t\rightarrow \infty }P_i(t)$, independent of
the initial conditions. This implies that $P_i^{*}$ is a \emph{right}
eigenvector of $W$ with eigenvalue zero. This eigenvalue is non-degenerate
so that $P_i^{*}$ spans the null space of the matrix $W_i^j$. The associated 
\emph{stationary currents} are denoted by $K^{*}{}_i^j$. They satisfy the
equality $\sum_{j\neq i}K^{*}{}_i^j=0$, for all $i$, i.e., the total
probability current into any given configuration vanishes.

In general, $P_i^{*}$ depends on the chosen rates, $w_i^j$. When simulating
systems in thermal equilibrium, the challenge is to specify a set $\{w_i^j\}$
such that the resulting stationary state equals the desired equilibrium
distribution, $P_i^{eq}$. A well-established procedure is to choose rates
which satisfy ``detailed balance'' (with respect to $P_i^{eq}$), namely, 
\begin{equation}
w_i^jP_j^{eq}-w_j^iP_i^{eq}=0  \label{db}
\end{equation}
for \emph{every pair} $\mathcal{C}_i$, $\mathcal{C}_j$ of configurations.
This relation can be viewed as a constraint on the set of allowable $w$'s.
We see immediately that Equation (\ref{db}) is equivalent to demanding that \emph{%
all individual currents vanish}, i.e., $(K^{eq}{})_i^j=0$ for all $i\neq j$.
Conversely, if the dynamics satisfies detailed balance, even if the 
steady-state distribution is not explicitly known, we can construct $P_i^{eq}$
easily from Equation (\ref{db}). Furthermore, it is now very easy to determine
whether two \emph{different }sets of rates, $\{w_i^j\}$ and $\{\bar{w}_i^j\}$%
, will generate the same $P_j^{eq}$: They do if the equalities $w_i^j/w_j^i=%
\bar{w}_i^j/\bar{w}_j^i$ hold.

Equation (\ref{db}) seems to imply that $P_{i}^{eq}$ must be explicitly known in
order to test the validity of detailed balance. However, detailed balance is
an intrinsic property of the dynamics, expressing a deeper statement on
microscopic reversibility and requiring no information about any specific
steady-state distribution $P_{i}^{eq}$. Known as the Kolmogorov criterion %
\cite{Kolmogorov,Kelly,DMrev,RBthesis}, it relies on considering closed
loops in configuration space, e.g., $\mathcal{L}$ $\equiv \mathcal{C}%
_{i}\rightarrow \mathcal{C}_{j}\rightarrow \mathcal{C}_{k}\rightarrow
...\rightarrow \mathcal{C}_{n}\rightarrow \mathcal{C}_{i}$. For each loop,
we define the product of the associated rates in the ``forward'' direction, $%
\Pi \left[ \mathcal{L}\right] \equiv w_{j}^{i}w_{k}^{j}\,...\,w_{i}^{n}$ , as
well as for the ``reverse'' direction: $\Pi \left[ \mathcal{L}_{rev}\right]
\equiv w_{i}^{j}w_{j}^{k}\,...\,w_{n}^{i}$. In terms of these products, the
dynamics is said to satisfy detailed balance if 
\begin{equation}
\Pi \left[ \mathcal{L}\right] =\Pi \left[ \mathcal{L}_{rev}\right] \quad 
\label{loops}
\end{equation}%
for \emph{all} loops.
This condition implies the \emph{path-independence} of the ratio of the
associated products of the rates along \emph{any} path which goes from $%
\mathcal{C}_{i}$ to $\mathcal{C}_{j}$. To be more explicit, consider a path: 
$\mathcal{P}$ $\equiv \mathcal{C}_{i}\rightarrow \mathcal{C}%
_{k}\,\,...\rightarrow \mathcal{C}_{n}\rightarrow \mathcal{C}_{j}$ and its
``reverse'' $\mathcal{P}_{rev}\equiv \mathcal{C}_{j}\rightarrow \mathcal{C}%
_{n}\,\,...\rightarrow \mathcal{C}_{k}\rightarrow \mathcal{C}_{i}$. Then, the
ratio 
\begin{equation}
\left. \Pi _{j}^{i}\left[ \mathcal{P}\right] \right/ \Pi _{i}^{j}\left[ 
\mathcal{P}_{rev}\right] \equiv \frac{w_{k}^{i}...w_{j}^{n}}{%
w_{n}^{j}...w_{i}^{k}}  \label{ratio}
\end{equation}%
assumes the same value as $\left. \Pi _{j}^{i}\left[ \mathcal{P}^{\prime }%
\right] \right/ \Pi _{i}^{j}\left[ \mathcal{P}_{rev}^{\prime }\right] $
where $\mathcal{P}^{\prime }$ is any other path from $\mathcal{C}_{i}$ to $%
\mathcal{C}_{j}$. Such path independence allows us to define a unique
``potential'' associated with each $\mathcal{C}_{i}$: 
\begin{equation}
\Phi _{io}\equiv \ln \left( \Pi _{i}^{o}/\,\Pi _{o}^{i}\right) \,\,,
\label{phi}
\end{equation}%
where $\mathcal{C}_{o}$ is some arbitrary reference configuration. As we
will demonstrate below, it is straightforward to
show that the stationary distribution is then given by 
\begin{equation}
P_{i}^{\ast }\propto \exp \left[ \Phi _{io}\right] \,\,.  \label{P_eq}
\end{equation}%
The relationship between this $P_{i}^{\ast }$ and the familiar $P_{i}^{eq}$
is now clear: The potential $\Phi $ is, e.g., just $-\beta \left\{ \mathcal{H%
}\left[ \mathcal{C}_{i}\right] -\mathcal{H}\left[ \mathcal{C}_{o}\right]
\right\} $ for the canonical distribution. Further, for systems obeying Equation (%
\ref{loops}), this approach allows us to define an effective Hamiltonian $%
\mathcal{H}_{eff}$ along with an effective (inverse) temperature $\beta
_{eff}$, in case such concepts are beneficial for the problem at hand.


A key signature of \emph{non}-equilibrium steady states is that the
underlying rates do not satisfy microscopic reversibility and Equation (\ref
{loops}) is violated. As a result, the ratios Equation (\ref{ratio}) will be
depend explicitly on the path $\mathcal{P}$. This leads Lebowitz and Spohn
to define \cite{LS-action} an action associated with each $\mathcal{P}$ as
the log of the ratio (\ref{ratio}).
In a more specific arena, the dynamic functional \cite{Janssen+} is its
counterpart in typical statistical field theories.

\begin{figure}[tbp]
\begin{center}
\epsfig{file=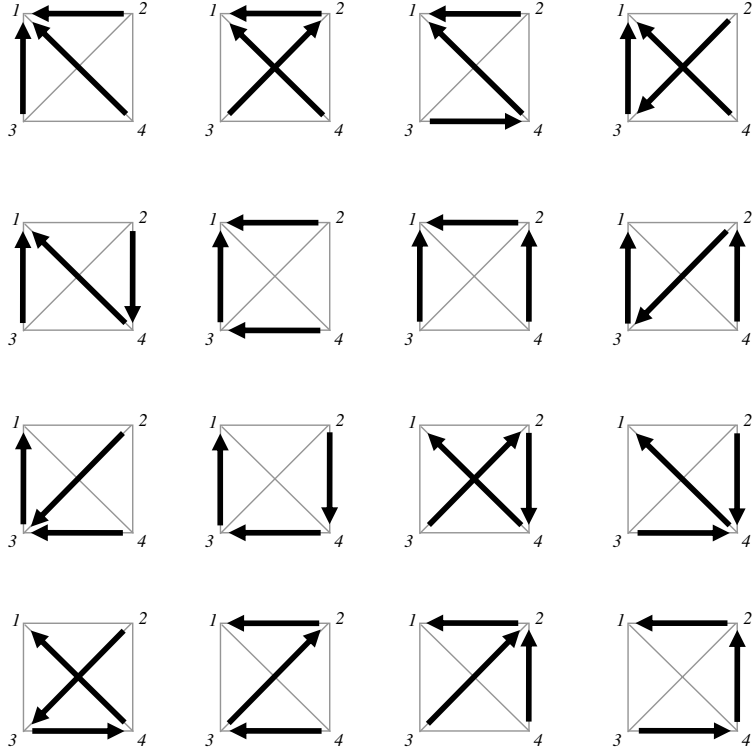,width=4.0in}
\end{center}
\par
\vspace{-0.4cm}
\caption{All 16 directed trees contributing to the graphical 
representation of $P^*_{1}$, for a simple model with $N=4$.}
\label{fig-1}
\end{figure}

For a NESS, the right hand side of Equation (\ref{db}) is generically nonzero,
given by the nontrivial stationary probability currents: As a result, the
construction of the stationary probability distribution -- while still
possible -- requires a bit more effort. Though established some time ago %
\cite{Hill,Schn,Haken}, the graph-theoretical approach, similar to those
originally designed for electric networks \cite{Kirchhoff}, appears not to
be widely known \cite{RBthesis}. For the reader's convenience, we briefly
review the method here. First, we represent each configuration by a labelled
vertex (e.g., $i$ for $\mathcal{C}_{i}$). Next, we draw all distinct
spanning trees (i.e. all distinct graphs containing all vertices and exactly
one single undirected edge between each pair so that no loops are formed) $%
t_{\alpha }$, $\alpha =1,2,..,M$, with $N$ vertices. The total number of
such trees, $M$, is given by $N^{N-2}$, according to Cayley's theorem \cite%
{Cayley}. Next, we select a specific vertex, e.g., $i$, and draw an arrow,
directed towards $i$, on every edge. The resulting \emph{directed} tree will
be denoted by $t_{\alpha (i)}$. Note that, for given $i$, there is exactly
one directed tree for each undirected tree. Next, we associate a factor $%
w_{j}^{k}$ with an edge directed from vertex $k$ to $j$. Finally, the
numerical value $U(t_{\alpha (i)})$ of the directed tree $t_{\alpha (i)}$ is
defined to be the product of the $(N-1)$ rates $w$ appearing in that
particular tree. If one of the rates vanishes, we simply assign the value $%
U(t_{\alpha (i)})=0$ to the associated tree. We illustrate the procedure in 
Figure~\ref{fig-1} for finding $P_{1}^{\ast}$ in a 
$N=4$ case. While we show all 16 directed
trees, we just give two of the $U$'s as examples: $%
U(t_{4(1)})=w_{1}^{4}w_{1}^{3}w_{3}^{2}$ (top right diagram in 
Figure~\ref{fig-1}) and 
$U(t_{13(1)})=w_{1}^{4}w_{4}^{3}w_{3}^{2}$ (bottom left diagram in 
\ref{fig-1}). The
stationary solution of Equation (\ref{me}) is then given by 
\begin{equation}
P_{i}^{\ast }=\mathcal{Z}^{-1}\sum_{\alpha }U(t_{\alpha (i)})  \label{trees}
\end{equation}%
with the normalization factor $\mathcal{Z}$ defined as 
\begin{equation}
\mathcal{Z}\equiv \sum_{i=1}^{N}\sum_{\alpha }U(t_{\alpha (i)})  \label{norm}
\end{equation}

Now that we have a representation for $P_{i}^{\ast}$, let us consider the
probability currents, defined in Equation (\ref{P-current}) above. Specializing
to the stationary case, we write the \emph{net }current from $\mathcal{C}%
_{j} $ into $\mathcal{C}_{i}$ in the form:

\begin{equation}
K^{*}{}_i^j\equiv w_i^jP_j^{*}-w_j^iP_i^{*}=\mathcal{Z}^{-1}\sum_\alpha %
\left[ w_i^jU(t_{\alpha (j)})-w_j^iU(t_{\alpha (i)})\right]
\label{st-current}
\end{equation}

\begin{figure}[tbp]
\begin{center}
\epsfig{file=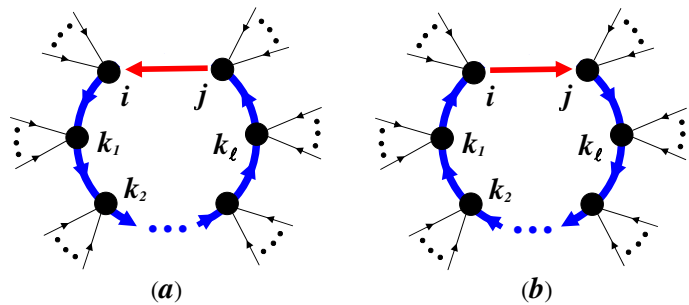,width=4.0in}
\end{center}
\par
\vspace{-0.4cm}
\caption{An illustration of the forward and backward loops, appearing in 
the sum on the right hand side 
of Equation~(\ref{st-current}). See text for details.}
\label{fig-2}
\end{figure}

Focusing on the expression within $[...]$, we note that, for a specific $%
\alpha $, the trees $t_{\alpha (i)}$ and $t_{\alpha (j)}$ differ only in the
directed edges that connect vertices $i$ and $j$. In Figure~\ref{fig-2}, we illustrate
this statement with a tree that has $k_{1},...,k_{\ell }$ as the vertices 
between $i$ to $j$. In Figure~\ref{fig-2}a, $t_{\alpha (j)}$ involves a directed tree
used for $U(t_{\alpha (j)})$, with directed edges being blue. Similarly, in
Figure~\ref{fig-2}b, we have the same tree being used for $U(t_{\alpha (i)})$, the only
difference being the edges \emph{between }$i$ and $j$ which are now
reversed. Considering just this segment of the two trees, we may write the
associated products of the $w$'s as $\Pi _{j}^{i}(t_{\alpha
(j)})=w_{k_{1}}^{i}w_{k_{2}}^{k_{1}}...w_{j}^{k_{\ell }}$ and $\Pi
_{i}^{j}(t_{\alpha (i)})=w_{k_{\ell }}^{j}...w_{i}^{k_{1}}$, respectively.
Meanwhile, the rest of both trees (the side branches ``coming into the
circle'' in Figure~\ref{fig-2}) are identical. Thus, denoting the products of the rates
associated with these side branches as $R$, we may write 
\begin{equation*}
R(t_{\alpha (i)})=R(t_{\alpha (j)}).  
\end{equation*}%
Combining the side branches with the path between $i$ and $j$, we arrive at 
\begin{eqnarray}
U(t_{\alpha (j)}) =\Pi_{j}^{i}(t_{\alpha (j)})R(t_{\alpha (j)})
\quad {\rm and} \quad
U(t_{\alpha (i)}) =\Pi_{i}^{j}(t_{\alpha (i)})R(t_{\alpha (i)})
\label{UPR}
\end{eqnarray}%
Returning to the steady-state current, Equation (\ref{st-current}), we see that
the additional factor of $w$ in each term can be regarded as adding an extra
edge (red), so that each tree is converted into a graph with a \emph{single }%
loop (the ``circles'' in Figure~\ref{fig-2}). It is natural to label these directed
loops as $\mathcal{L}_{\alpha \left( j\right) }$ and $\mathcal{L}_{\alpha
\left( i\right) }$ respectively. Of course, both refer to the \emph{same}
loop, just traversed in \emph{opposite} directions. Quantitatively, we have,
for each $t_{\alpha }$%
\begin{equation*}
w_{i}^{j}U(t_{\alpha (j)})-w_{j}^{i}U(t_{\alpha (i)})=\left\{
w_{i}^{j}w_{k_{1}}^{i}w_{k_{2}}^{k_{1}}...w_{j}^{k_{\ell
}}-w_{j}^{i}w_{k_{\ell }}^{j}...w_{i}^{k_{1}}\right\} R(t_{\alpha (j)})\,\,.
\end{equation*}%
Now, the terms in $\left\{ ...\right\} $ can be associated with $\mathcal{L}%
_{\alpha \left( j\right) }$ and $\mathcal{L}_{\alpha \left( i\right) }$, so
that a compact expression for the current is 
\begin{equation}
K^{\ast }{}_{i}^{j}=\mathcal{Z}^{-1}\sum_{\alpha }\left[ \Pi (\mathcal{L}%
_{\alpha (j)})-\Pi (\mathcal{L}_{\alpha (i)})\right] R(t_{\alpha (i)})\,\,.
\label{current-loops}
\end{equation}%
From here, it is immediately obvious that detailed balance, manifested in
reversible loops, Equation (\ref{loops}), will lead to all $K^{\ast }{}_{i}^{j}$
being zero, and nontrivial steady-state currents can only emerge from rates
which violate detailed balance. For completeness, we should remark that it
is conceivable for accidental cancellations in the sum over trees to occur
so that \emph{some} $K^{\ast }$'s may vanish even though irreversible loops are
present. However, $K^{\ast }$ should be nonzero, generically.

Let us point out briefly that there is an alternate representation of the
stationary distribution \cite{RBthesis}, in terms of the
co-factors of $W_{i}^{j}$ (which we denote by $C_{j}^{i}$ here). Of course,
a well known expression of the determinant leads us to $\det
W=\sum_{j}W_{i}^{j}C_{j}^{i}$ for any $i$. Moreover, 
for a stochastic matrix $W_{i}^{j}$, $C_{j}^{i}$ is independent of $i$. 
Thus, we
may write $C_{j}$ in place of $C_{j}^{i}$. Finally, since $W$ has a zero
eigenvalue, we have $0=\det W=\sum_{j}W_{i}^{j}C_{j}$. Thanks to the
uniqueness of the stationary solution, this equality implies $C_{j}\propto
P_{j}^{\ast }$.

Before proceeding to the next section, it may be instructive to see how the
complex expression for $P^{\ast }$, Equation (\ref{trees}), reduces directly to the
familiar form for systems in equilibrium. To start, we choose an arbitrary
reference configuration $\mathcal{C}_{o}$. Of course, we will rely on the 
path-independent properties of $\Pi _{j}^{o}/\Pi _{o}^{j}$. Consider the ratio 
\begin{equation*}
\left. 
\sum_{\alpha }U(t_{\alpha (j)}) \left[ \sum_{\alpha }U(t_{\alpha (o)})\right]^{-1}
\right. \end{equation*}%
and exploit the factorization, Equation (\ref{UPR}). Then, the numerator of the
above reads 
\begin{equation}
\sum_{\alpha }U(t_{\alpha (j)})=\sum_{\alpha
}\Pi_{j}^{o}(t_{\alpha (j)})R(t_{\alpha (j)})\,\,,  \label{UPRo}
\end{equation}%
where $\Pi _{j}^{o}(t_{\alpha (j)})$ now stands for the product of the rates
associated with the path from $\mathcal{C}_{o}$ to $\mathcal{C}_{j}$ in $%
t_{\alpha (j)}$ and $R(t_{\alpha (j)})$, with the rest of the tree (``side
branches''). Of course, we have a similar decomposition for the denominator. 
Next, we invoke the path independence of the 
$\Pi $'s (Equation (\ref{phi})) and write 
\begin{equation*}
\Pi_{j}^{o}(t_{\alpha (j)})=\Pi_{o}^{j}(t_{\alpha
(o)})e^{\Phi _{jo}}\,\,.
\end{equation*}%
Let us emphasize that $\Phi _{jo}$ does \emph{not} depend on the path so
that it is also independent of $t_{\alpha }$. Consequently, Equation (\ref{UPRo})
becomes 
\begin{equation*}
\sum_{\alpha }U(t_{\alpha (j)})=e^{\Phi _{jo}}\sum_{\alpha
}\Pi_{o}^{j}(t_{\alpha (o)})R(t_{\alpha (j)})\,\,.
\end{equation*}%
But, as noted above, the side branches of $t_{\alpha (j)}$ and $t_{\alpha
(o)}$ are identical, so that we can replace $R(t_{\alpha (j)})$ by $%
R(t_{\alpha (o)})$. The sum on the right is easily recognized as $%
\sum_{\alpha }U(t_{\alpha (o)})$ and we arrive at 
\begin{equation*}
\sum_{\alpha }U(t_{\alpha (j)})=e^{\Phi _{jo}}\sum_{\alpha }U(t_{\alpha
(o)})\,\,.
\end{equation*}%
and the desired result for systems in thermal equilibrium: 
\begin{equation}
P_{i}^{\ast }=P_{o}^{\ast }\exp \left[ \Phi _{io}\right] \,\,.
\label{P_eq 1}
\end{equation}

\section{A postulate for the complete characterization of NESS}
\label{post}

In much of the literature on NESS, the \emph{microscopic} stationary
distribution $P^{\ast }$ tends to be the central focus, along with those 
\emph{macroscopic} properties which can -- at least in principle -- be
derived from it, such as order parameters and correlation functions. In this
sense, the investigations follow standard routes for systems in thermal
equilibrium. Indeed, many studies emphasize the similarity between the $%
P^{\ast }$ of a NESS and the $P^{eq}$ of an equilibrium system. At the same
time, many complementary studies focus on quantities \emph{absent} from
equilibrium systems, e.g., particle currents or energy fluxes through the
system. Unlike $P^{\ast }$, however, these are \emph{macroscopic} averages,
and thus located at the same scale as order parameters and correlation
functions. It is natural to ask whether there is a non-equilibrium quantity,
or concept, which would provide a \emph{microscopic }basis for these fluxes.
Here, we propose that the microscopic distribution of probability currents, $%
K^{\ast }$, be brought to stage center, as an indispensable partner for
the microscopic probability distribution $P^{\ast }$. Clearly, the
particle/energy fluxes can be computed from $K^{\ast }$ as averages, on a
par with $\left\langle \bullet \right\rangle =\Sigma \bullet P^{\ast }$.
Needless to say, in the arena of \emph{equilibrium} statistical physics, all 
$K^{\ast }$'s vanish and therefore play the role of an invisible partner.

To showcase the essential nature of $K^{\ast }$, let us consider a
completely trivial example, namely, a single particle hopping randomly
between neighboring sites on a ring. As long as the rates are spatially
uniform, the stationary probability distribution is trivially flat: $P^{\ast
}\propto 1$, \emph{regardless }of whether the hopping rates are symmetric
(i.e., identical for left and right hops) or biased. 
Yet, detailed balance is satisfied
if and only if they are symmetric, and there is an important \emph{physical}
difference between the two cases, namely, the presence or absence of a
particle current in the stationary state. In more complex cases, other types
of currents (e.g., energy) may be the key feature. In general, all these
fluxes can be traced to the violation of detailed balance and the associated
microscopic probability currents. In other words, without $K^{\ast }$, $%
P^{\ast }$ alone cannot be a complete description of a NESS.

That $K^{\ast }$ has been largely ignored in the literature may be traced to
the following. In all studies of non-equilibrium systems, we begin with a
given set of rates $\left\{ w\right\} $, motivated by the underlying
physics, chemistry, biology, psychology, sociology, etc. The main difficulty
is to find $P^{\ast }$. But, once $P^{\ast }$ is known, macroscopic average
fluxes can be computed easily from $w$ and $P^{\ast }$, so that it is
unnecessary to construct $K^{\ast }$ explicitly. However, if we wish to ask
the \emph{inverse question}, namely: ``What rates are needed to \emph{achieve%
} a particular NESS?'', then $K^{\ast }$ plays an indispensable role.

Though seemingly trivial, let us state for completeness that $K^{\ast }$
alone is also inadequate. Specifically, it is possible to have NESS's with
the same $K^{\ast }$ but different $P^{\ast }$'s. Electromagnetism provides
a good analog. In electrostatics, the central focus is the charge
distribution, being the analog of $P^{\ast }$. By definition of
electrostatics, there are no currents anywhere. Similarly, we are concerned
mainly with the currents in magnetostatics and tend to ignore the charge
distribution. However, it is clear that, in general, the charge distribution
is not trivial, especially for non-neutral systems.

Staying with electromagnetism for a moment, Kirchhoff's solution for general
electric circuits \cite{Kirchhoff} is often quoted as the first graphical
solution to the master equation. Yet, there are non-trivial differences.
Specifically, the currents ($I_{ij}$ between nodes $i$ and $j$) are clearly
the quantities of interest for Kirchhoff, while there is no trace of the
charge distribution ($\rho _{i}$ at node $i$) in his solution. In contrast, $%
P^{\ast }$ is the key quantity when solving the master equation. Further,
there is no one-to-one mapping from the set of electromotive forces and
resistances to the set of rates. Further details regarding the ``duality''
of our NESS and the Kirchhoff problem may be found in Appendix A.

Motivated by these considerations, we propose that $\{P^{*},K^{*}\}$, the
distributions for probability and probability currents, form a \emph{%
complete and unique} description for\emph{\ any }stationary state. This
classification includes equilibrium systems, characterized by $\{P^{eq},0\}$%
, as well as NESS's where $K^{*}$ is, by definition, non-zero.

The inclusion of $K^{\ast }$ in this characterization motivates the
definition of the ``distance'' of a given NESS\ from equilibrium, namely, 
\begin{equation}
\left| K^{\ast }\right| ^{2}\equiv \sum_{i,j}(K^{\ast }{}_{j}^{i})^{2}\,\,.
\label{K^2}
\end{equation}%
This may provide a quantitative basis for phrases like ``near equilibrium''
and ``far from equilibrium.'' How useful this concept is remains to be
explored. In Section \ref{entropy} below, we will see $(K^{\ast }{}_{j}^{i})^{2}$
appearing in a discussion of entropy production.

More importantly, $K^{\ast }$ contains all information necessary to
determine the average \emph{fluxes }(or currents) associated with physical
observables, such as energy or particle number density. A microscopic
distribution of probability currents, it serves as the statistical weight in
the computation of physical currents. Explicitly, we write 
\begin{equation}
\left\langle \mathcal{J}\right\rangle \equiv \frac{1}{2}\sum_{i,j}\mathcal{J}%
\left( \mathcal{C}_{i},\mathcal{C}_{j}\right) K^{\ast }\left( \mathcal{C}%
_{i},\mathcal{C}_{j}\right)  \label{fluxes J}
\end{equation}%
where $\mathcal{J}\left( \mathcal{C}_{i},\mathcal{C}_{j}\right) $ is
associated with a physical observable. Naturally, we expect physically
meaningful $\mathcal{J}$'s to be antisymmetric under $\mathcal{C}%
_{i}\Leftrightarrow \mathcal{C}_{j}$ (e.g., a particle current). Taking
advantage of this fact, we can define $\left\langle \mathcal{J}\right\rangle 
$ in a manner that highlights $K^{\ast }$ as a probabilistic weight, namely, 
\begin{equation}
\left\langle \mathcal{J}\right\rangle \equiv \sum_{\left\{ i,j\right\} }%
\mathcal{J}\left( \mathcal{C}_{i},\mathcal{C}_{j}\right) K^{\ast }\left( 
\mathcal{C}_{i},\mathcal{C}_{j}\right)  \label{{fluxes J}}
\end{equation}%
where $\left\{ i,j\right\} $ denotes a sum taken over \emph{only the positive%
} $K^{\ast }$s. For the readers' convenience, we illustrate these
considerations with two examples.

Many NESS models involve particles hopping from a site (labeled by $s$) to a
neighboring site on a lattice, with an excluded volume constraint. A classic
example is the totally asymmetric exclusion process (TASEP) \cite%
{earlyTASEP,Schuetz}. For such models, a configuration is uniquely given by
the set of occupation numbers: $\left\{ n_{s}\right\} $. Now, one quantity
of interest might be the average particle current across a \emph{particular}
pair of sites (bond). Specifically, for the current from site $a$ to $b$, we
need 
\begin{equation}
\mathcal{J}_{bond}\left( \left\{ n_{s}\right\} ,\left\{ n_{s}^{\prime
}\right\} \right) =n_{a}\left( 1-n_{b}\right) \left( 1-n_{a}^{\prime
}\right) n_{b}^{\prime }-n_{a}^{\prime }\left( 1-n_{b}^{\prime }\right)
\left( 1-n_{a}\right) n_{b}\,  \label{Jbond}
\end{equation}%
Clearly, for the net current across, say, some surface in a 3-dimensional
system, we only need to sum the $\mathcal{J}$'s associated with bonds which
pierce the surface: $\mathcal{J}_{surface}\left( \left\{ n\right\} ,\left\{
n^{\prime }\right\} \right) =\Sigma \mathcal{J}_{bond}$. Further, if the
entire distribution for this current (denoted by $p\left( \mathcal{J}\right) 
$) is of interest, then it is given by 
\begin{equation*}
p\left( \mathcal{J}\right) =\frac{1}{2}\sum_{\left\{ n\right\} ,\left\{
n^{\prime }\right\} }\delta \left( \mathcal{J}-\mathcal{J}_{surface}\left(
\left\{ n\right\} ,\left\{ n^{\prime }\right\} \right) \right) K^{\ast
}\left( \left\{ n\right\} ,\left\{ n^{\prime }\right\} \right) \,\,.
\end{equation*}

Let us reiterate the need for $K^{\ast }$ (as opposed to $P^{\ast }$) in
finding these physical currents, even though $P^{\ast }$ alone may appear to
be the only necessary ingredient for computing averages. This paradox can be
traced to Equation (\ref{P-current}), which allows us access to $K^{\ast }$ from $%
P^{\ast }$, provided the rates $\left\{ w\right\} $ are known. Typical
studies of NESS begin with a given set of rates and unknown $\left\{ P^{\ast
},K^{\ast }\right\} $. Thus, the currents are seemingly unimportant, once $%
P^{\ast }$ is found. However, as will be discussed in the next section,
there are many $w$'s that can lead to the same $\left\{ P^{\ast },K^{\ast
}\right\} $! In this sense, the rates are not \emph{necessary }(though
sufficient) for finding average fluxes, and some details of their precise
form are ``irrelevant.''

\begin{figure}[tbp]
\begin{center}
\epsfig{file=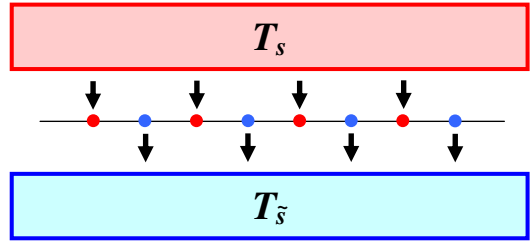,width=3.0in}
\end{center}
\par
\vspace{-0.4cm}
\caption{A simple kinetic Ising chain, coupled to two temperature baths. Spins coupled to the
higher (lower) temperature are indicated by red (blue) dots. In the steady state, 
there are constant energy fluxes, as denoted by
the black arrows. }
\label{fig-3}
\end{figure}

Another example of a current in a simple NESS is the total energy flux
through a kinetic Ising model coupled to two thermal baths at \emph{%
different }temperatures, as studied in, e.g., \cite%
{GM-TTIS,NL+me-TTIS,ZRRZ,SZrev,SS1,SS2}. For simplicity, we consider
a one-dimensional version here, as shown in Figure \ref{fig-3}. 
Let us denote the set of sites
coupled to the two baths by $\left\{ s\right\} $ and $\left\{ \tilde{s}%
\right\} $, respectively, and a particular spin configuration by $\left\{
\sigma _{s}\right\} $. We are interested in the energy flowing from 
the bath at the higher temperature, $T_{s}$, to the bath at lower 
temperature, $T_{\tilde{s}}$. Given the details of the coupling,
energy flows into (out of) the system through the $\left\{s \right\} $ 
($\left\{ \tilde{s} \right\}$)
spins. Using a random sequential dynamics, the energy current operator, 
in units of the nearest-neighbor coupling strength, can be written as
\begin{eqnarray*}
\mathcal{J_{E}} \left( \left\{ \sigma _{s}\right\} ,\left\{ \sigma _{s}^{\prime
}\right\} \right) &=&\sum_{s}\left( \sigma _{s-1}+\sigma _{s+1}\right) \left(
\sigma _{s}^{\prime }-\sigma _{s}\right) \delta _{s} \nonumber \\
&-&\sum_{\tilde{s}}\left(
\sigma _{\tilde{s}-1}+\sigma _{\tilde{s}+1}\right) \left( \sigma _{\tilde{s}%
}^{\prime }-\sigma _{\tilde{s}}\right) \delta _{\tilde{s}}\,\,,
\end{eqnarray*}%
where $\delta _{s}\equiv \prod_{k\neq s}\left( 1+\sigma _{k}\sigma
_{k}^{\prime }\right) /2$ and a similar $\delta _{\tilde{s}}\,$ insure that
only the spin at site $s$ or $\tilde{s}$ is flipped. 
If the two temperatures are the same and the
system is in equilibrium, then $\left\langle \mathcal{J_{E}} \right\rangle $ is
trivially zero due to $K^{\ast }\equiv 0$. If we use the rates and $P^{\ast
} $ instead, a tedious calculation will show that the average of \emph{each
term} in $\mathcal{J_{E}} $ vanishes.

To summarize, we postulate that $\{P^{\ast },K^{\ast }\}$ is a complete
characterization of a general NESS. Compared to an equilibrium state, which
is specified by $N-1$ quantities ($P^{\ast }$), a NESS requires $O\left(
N^{2}\right) $\ quantities. Note that the $N\left( N-1\right) /2$ quantities
in $K^{\ast }$ are not independent: There are $N$ constraints, namely, $%
\Sigma _{i}K^{\ast }{}_{j}^{i}=0$. The significance of our characterization
is that $\{P^{\ast },K^{\ast }\}$ gives access to all macroscopic quantities
of interest, including any currents flowing through, or current loops
within, the system, without requiring any additional knowledge about the
rates. In the following, we explore the consequences of our proposal.

\section{Consequences of the postulate}
\label{cons}

Once a NESS is characterized by $\{P^{\ast },K^{\ast }\}$, we can explore
the possibility of a generalized ``detailed balance condition,'' which the
transition rates must satisfy in order to ensure settling into a given NESS.
This freedom of choice has implications for entropy production, as well as
for the design of optimized computer simulations.

\subsection{Dynamic equivalence classes.}
\label{DEC}

In many investigations of NESS, one of the most striking features is their
sensitivity to the details of the transition rates. Major changes of
macroscopic properties often emerge from seemingly minor modifications of
the rates. It is therefore natural to ask how one might determine whether
two sets of rates will lead to the same, or a different, NESS. This issue
can be addressed within the framework proposed here. In particular, since
all properties of a NESS are supposedly given by $\{P^{\ast },K^{\ast }\}$,
all \emph{time-independent }(static) quantities of physical interest can be
computed, \emph{without} further recourse to the dynamics 
\cite{notall}. The analog in equilibrium
systems is that all (static) quantities can be obtained from $P^{\ast }$
without detailed knowledge of the rates.

An alternate phrasing of the above question is: Given a set of $w$'s and its
associated NESS, what are the transformations on the rates which leave this $%
\{P^{\ast },K^{\ast }\}$ invariant? For the equilibrium case, $\{P^{eq},0\}$%
, the answer is provided by the detailed balance condition, Equation (\ref{db}):\
Any set of $w$'s will lead to the desired $P^{eq}$, provided $%
w_{i}^{j}/w_{j}^{i}=P_{i}^{eq}/P_{j}^{eq}$. Alternatively, this relation can
be interpreted as a constraint to be placed on the rates if the goal is to
reach a specific $P^{eq}$. In our framework, this constraint can now be
easily generalized: To arrive at a \emph{given }$\{P^{\ast },K^{\ast }\}$
final state, the $w$'s must satisfy \ 
\begin{equation}
w_{i}^{j}P_{j}^{\ast }-w_{j}^{i}P_{i}^{\ast }=K^{\ast }{}_{i}^{j}\,\,.
\label{w-rel}
\end{equation}%
for all pairs $i\neq j$. As a simple extension of the equilibrium case (Equation %
\ref{db}), this equation constrains, say, the ``backward'' rate $w_{j}^{i}$
once an (arbitrary or suitable) ``forward'' rate $w_{i}^{j}$ is selected. In
this sense, there is just as much freedom in choosing rates to arrive at a
given NESS as for systems to reach thermal equilibrium. In the following, we
will explore alternative expressions for this simple constraint, to provide
further insight.

Motivated by the antisymmetric nature of the currents $K^{*}{}_i^j$, we
define 
\begin{equation}
\bar{W}_i^j\equiv W_i^jP_j^{*}  \label{W-bar}
\end{equation}
and decompose it into its symmetric and antisymmetric parts:\ 
\begin{equation}
\bar{W}_i^j=S_i^j+A_i^j\,\,,  \label{def-M}
\end{equation}
where $S_i^j\equiv (\bar{W}_i^j+\bar{W}_j^i)/2$, and $A_i^j\equiv (\bar{W}%
_i^j-\bar{W}_j^i)/2$. Then, Equation (\ref{w-rel}) is a very simple constraint,
namely, 
\begin{equation}
A_i^j=\frac 12K^{*}{}_i^j\,\,.  \label{fix-A}
\end{equation}
As a result, for a given $\{P^{*},K^{*}\}$, $A$ is completely determined. In
contrast, $S$ is essentially free, except for two restrictions. First, the
physical rates must be non-negative ($w\geq 0$), leading to $S_i^j\geq
|A_i^j|$, for all $i\neq j$. Further, probability conservation imposes $%
\sum_iW_i^j=0$, for all $j$, so that $\sum_i\bar{W}_i^j=0$ also. Now, $%
\sum_iA_i^j$ also vanishes, since $\sum_i(\bar{W}_i^j-\bar{W}%
_j^i)=-\sum_iW_j^iP_i^{*}$, which is zero by virtue of 
$\partial _tP_{j}^{\ast}=0$.
Therefore, $\sum_i\bar{W}_i^j=0$ reduces to $\sum_iS_i^j=0$ and we arrive at
the restrictions for $S$: 
\begin{equation}
S_i^j\geq \frac 12\left| K^{*}{}_i^j\right| \quad \forall i\neq j,\quad
\quad S_j^j=-\sum_{i\neq j}S_i^j\,.  \label{M-const}
\end{equation}
Within these constraints, we may choose \emph{arbitrary} $S$'s, construct
transition rates via 
\begin{equation}
W_i^j=\left[ S_i^j+\frac 12K^{*}{}_i^j\right] (P_j^{*})^{-1}\quad ,
\label{new-W}
\end{equation}
and be certain that the final NESS remains unaffected. In this respect, we
may regard all such $S$'s - specified by $N\left( N-1\right) /2$ parameters
- as generating an ``equivalence class'' of dynamic rates associated with
the \emph{same }NESS.

There is another, perhaps simpler, way to express this freedom of choice.
Suppose we found $P^{\ast }$ from a given set of rates $w_{i}^{j}$ (and so, $%
K^{\ast }$ is also determined). To construct another set of equivalent
rates, it is sufficient to \emph{add} to the $w$'s any set of changes $%
\Delta $ that satisfy 
\begin{equation}
\Delta _{i}^{j}P_{j}^{\ast }=\Delta _{j}^{i}P_{i}^{\ast }\,\,.
\label{Delta-db}
\end{equation}%
Note that $\Delta $ may be\emph{\ negative}, provided the new rates $%
w_{i}^{j}+\Delta _{i}^{j}$ are non-negative. Of course, this amounts to the
same statement as Equation (\ref{M-const}). Equation (\ref{Delta-db}) is reminiscent
of the ``ordinary detailed balance'' condition. The distinguishing feature
here is that the \emph{differences }between two sets of rates, as opposed to
the rates themselves, must satisfy ``detailed balance with respect to $%
P^{\ast }$.''

Another interesting corollary allows us to determine (some aspects of) $P^{\ast }$
from \emph{two} different sets of rates $w$, $w^{\prime }$ provided we know
that both lead to the same NESS. The key is to compute the differences: $%
\Delta =w-w^{\prime }$. For all nonvanishing pairs $\left\{ \Delta
_{i}^{j},\Delta _{j}^{i}\right\} $, the ratio can be used to construct $%
P^{\ast }$. In this sense, the $\Delta $'s can be thought of as `rates that
lead to an \emph{equilibrium }state', satisfying microscopic reversibility,
Equation (\ref{loops}). Two features distinguish this case from a true
equilibrium system: (a) Some $\Delta $'s can be negative and (b) there is a
unique $P^{\ast }$ even if the $\Delta $'s are not ergodic (i.e., not all
configurations connected by $\Delta $'s).

Lastly, let us present an alternate approach. For equilibrium steady states,
instead of considering $\bar{W}$, it is more convenient to define another
matrix, namely, 
\begin{equation}
\tilde{W}_i^j\equiv W_i^j\left( P_j^{*}/P_i^{*}\right) ^{1/2}
\label{W-twid}
\end{equation}
Due to detailed balance, $\tilde{W}$ is symmetric, and the master equation
is transformed into a standard eigenvalue problem. Apart from a single zero
eigenvalue (associated with $P^{eq}$), all other eigenvalues are negative.
Clearly, we can also consider the properties of $\tilde{W}$ in the
non-equilibrium case. Similar to $\bar{W}$, $\tilde{W}$ can be decomposed
into a symmetric ($\tilde{S}$) and a \emph{nonzero} asymmetric part, $%
\tilde{A}=\tilde{K}^{*}/2$. Here, $\tilde{K}^{*}$ is related to the original
currents $K^{*}$ via 
\begin{equation}
\tilde{K}^{*}{}_i^j\equiv K^{*}{}_i^j\left( P_j^{*}P_i^{*}\right) ^{-1/2}%
\qquad .  \label{K-twid}
\end{equation}
For the NESS, we can still conclude, thanks to the Perron-Frobenius theorem,
that there is only a single zero eigenvalue, and that all other eigenvalues
have strictly negative real parts. However, the eigenvalues need no longer
be real and may appear in complex conjugate pairs. These correspond to
oscillatory relaxation, as in underdamped oscillators.

Similar to our approach above, for a given NESS, $\tilde{A}$ is completely
determined while $\tilde{S}$ is essentially arbitrary (apart from the minor
restrictions imposed by positivity and normalizability). The two approaches
have complementary advantages: $\bar{W}$ provides immediate information on
the probability currents. $\tilde{W}$ has the same spectrum as the original $%
W$, and therefore reflects the dynamics more clearly.

We conclude this section by emphasizing that $W$ matrices belonging to an
equivalence class have, in general, a different set of nonzero eigenvalues.
Needless to say, the physical interpretation is that \emph{different} rates
generally lead to \emph{different} relaxation rates into the \emph{same} NESS.
Indeed, our hope is that future Monte Carlo simulations of the NESS will
exploit this freedom to devise more efficient algorithms, in the same spirit
as, say, the cluster algorithms \cite{Swendsen-Wang}, designed specifically
to circumvent critical slowing-down in equilibrium systems near critical
points.

\subsection{Entropy production }
\label{entropy}

One of the key signatures of non-equilibrium steady states, recognized many
decades ago \cite{GP,Schn,Sei}, is entropy production. Using the expression
for average fluxes above (Equation \ref{fluxes J}), one possible definition \cite%
{Schn} for the total entropy production is:\ 
\begin{equation}
\mathbf{\dot{S}}_{tot}(t)\equiv \frac{1}{2}\sum_{i,j}K{}_{i}^{j}(t)\ln \frac{%
W_{i}^{j}P_{j}(t)}{W_{j}^{i}P_{i}(t)}\,\,.  \label{S_tot}
\end{equation}%
If the master equation is interpreted in the language of chemical reactions, 
$\ln \left( W_{i}^{j}P_{j}/W_{j}^{i}P_{i}\right) $ would be an affinity, or
generalized ``thermodynamic force'' \cite{Schn}. Inserting Equation (\ref%
{P-current}) for $K{}_{i}^{j}(t)$, it is immediately apparent that $\mathbf{%
\dot{S}}_{tot}(t)\geq 0$. Further, $\mathbf{\dot{S}}_{tot}$ can be written
as the sum of two terms, i.e., the entropy production of the ``system'' and
of the ``medium'', 
\begin{equation}
\mathbf{\dot{S}}_{sys}(t)\equiv \frac{1}{2}\sum_{i,j}K{}_{i}^{j}(t)\ln \frac{%
P_{j}(t)}{P_{i}(t)}\,\,,\,\,\,\mathbf{\dot{S}}_{med}(t)\equiv \frac{1}{2}%
\sum_{i,j}K{}_{i}^{j}(t)\ln \frac{W_{i}^{j}}{W_{j}^{i}}\,.  \label{S-sys+med}
\end{equation}%
The former is readily recognized as the time derivative of $\mathbf{S}%
_{sys}\equiv -\sum_{i}P_{i}(t)\ln P_{i}(t)$, which motivates the term
``entropy production of the system''. The latter is attributed to the
coupling of the system to the \emph{external environment} in a manner that
prevents it from reaching equilibrium \cite{Schn}. Unlike $\mathbf{\dot{S}}%
_{tot}$, neither $\mathbf{\dot{S}}_{sys}$ nor $\mathbf{\dot{S}}_{med}$ are
necessarily positive.

Since we are primarily interested in steady states, we take the infinite
time limit. For equilibrium systems, all $K^{*}{}_i^j$ are identically zero
so that both $\mathbf{\dot{S}}_{sys}$ and $\mathbf{\dot{S}}_{med}$ vanish.
For a NESS, however, only $\mathbf{\dot{S}}_{sys}^{*}$ vanishes. In general, 
$K^{*}\neq 0$ leads to $\mathbf{\dot{S}}_{med}^{*}=\mathbf{\dot{S}}%
_{tot}^{*}>0$. The interpretation of these results is clear: In the steady
state, the entropy associated with the system no longer changes. However,
being coupled in an irreversible way to the environment, a NESS continues to
increase the entropy of its surrounding medium. In other words, $\mathbf{%
\dot{S}}_{med}^{*}$ carries information about the precise nature of these
couplings, encoded in the transition rates. So, even if we insist on having
the same NESS (i.e., a given $\{P^{*},K^{*}\}$), $\mathbf{\dot{S}}_{med}^{*}$
will not be unique. In the following, we explore the connection between $%
\mathbf{\dot{S}}_{med}^{*}$ and the transition rates in more detail.

In the steady state, $\mathbf{\dot{S}}_{tot}^{*}=\mathbf{\dot{S}}_{med}^{*}$%
. Recalling the decomposition of $WP^{*}$ into a symmetric and an
antisymmetric component, Equation (\ref{def-M}), it is more convenient to work
with $\mathbf{\dot{S}}_{tot}^{*}$ whence \ 
\begin{equation}
\mathbf{\dot{S}}_{tot}^{*}=\frac 12\sum_{i,j}K^{*}{}_i^j\ln \frac{%
W_i^jP_j^{*}}{W_j^iP_i^{*}}=\frac 12\sum_{i,j}K^{*}{}_i^j\ln \frac{%
S_i^j+A_i^j}{S_i^j-A_i^j}\,\,.  \label{S-tot-star}
\end{equation}
While the antisymmetric component is fixed by the currents, $A=K^{*}/2$, the
symmetric part $S$ can be chosen at will, as long as it satisfies Equation (\ref%
{M-const}). In particular, while $\mathbf{\dot{S}}_{tot}^{*}>0$ has to
remain valid for any choice of $S$, it is possible to make the entropy
production arbitrarily small or infinitely large. In other words, exploiting
the freedom in the choice of the transition rates, we can select an \emph{%
arbitrary} value of $\mathbf{\dot{S}}_{tot}^{*}$, without affecting the
underlying NESS\ $\{P^{*},K^{*}\}$. Let us provide a few details.

To achieve an arbitrarily small $\mathbf{\dot{S}}_{tot}^{\ast }$, we only
need $S\gg A$, for every non-vanishing element. Expanding Equation (\ref%
{S-tot-star})\ to lowest order in $A_{i}^{j}/S_{i}^{j}=K^{\ast
}{}_{i}^{j}/(2S_{i}^{j})$, we arrive at 
\begin{equation}
\mathbf{\dot{S}}_{tot}=\sum_{i,j}\frac{\left( K^{\ast }{}_{i}^{j}\right) ^{2}%
}{2S_{i}^{j}}\left[ 1+O\left( \frac{K^{\ast }}{S}\right) ^{2}\right]
\label{S-min}
\end{equation}%
Of course, even though $\mathbf{\dot{S}}_{tot}^{\ast }$ can be made
arbitrarily small, it still remains strictly positive, retaining the NESS
signature thanks to $\left( K^{\ast }{}_{i}^{j}\right) ^{2}$ being strictly
positive.
Since $\mathbf{\dot{S}}_{med}=\mathbf{\dot{S}}_{tot}=0$
characterizes an equilibrium system, minimizing $\mathbf{\dot{S}}%
_{tot}^{\ast }$ (for a given NESS)\ corresponds to selecting rates which are
as ``equilibrium-like'' as possible.

At the opposite extreme, we can construct rates with ``infinite'' $\mathbf{%
\dot{S}}_{med}^{\ast }$ by reducing at least one off-diagonal $S_{i}^{j}$ to 
$\left| A_{i}^{j}\right| $, so that either $S_{i}^{j}+A_{i}^{j}$ or $%
S_{i}^{j}-A_{i}^{j}$ vanishes (but never both). Translating this back into a
new matrix of transition rates, via Equation (\ref{new-W}), the corresponding
transition is now uni-directional, in that one of the two directed edges
between the associated pair of configurations is \emph{missing}. Of course,
it is possible to make all rates uni-directional, which may be naturally
referred to as ``maximally asymmetric.'' Such systems appear frequently in
the literature, a particularly familiar example being the totally asymmetric
exclusion processes (TASEP) \cite{earlyTASEP,Schuetz}. One clear advantage
of having maximally asymmetric rates for all edges is that the number of
nontrivial trees contributing to $P^{\ast }$ is kept at the absolute
minimum. Needless to say, the expression for $K^{\ast }$ also simplifies, to
just one term in Equation (\ref{st-current}), and all irreversible loops are
also uni-directional.

\begin{figure}[tbp]
\begin{center}
\epsfig{file=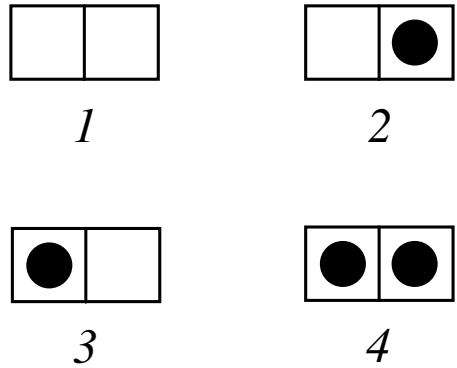,width=3.0in}
\end{center}
\par
\vspace{-0.4cm}
\caption{All configurations of a TASEP on a one-dimensional lattice with
just two sites. Particles are denoted by black circles.}
\label{fig-4}
\end{figure}

\section{Examples}
\label{examples}

The formalism we presented is quite general. It is instructive to provide a
series of simple examples, to illustrate how it applies in various
circumstances. We will start with a \emph{minute} system, a special case of
the $N=4$ example of \ref{master}. It is closely
associated with the zero range process (ZRP) and its generalizations which
we consider in the second subsection. The third subsection is devoted to a
kinetic Ising model coupled to two thermal baths at different
temperatures. Finally, we close with the general class of NESS with Gaussian
distributions.

\subsection{TASEP with two sites}
\label{examples-TASEP}

\begin{figure}[tbp]
\begin{center}
\epsfig{file=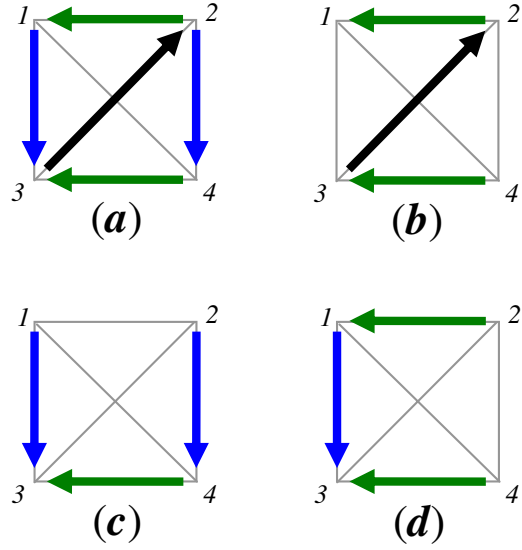,width=3.0in}
\end{center}
\par
\vspace{-0.4cm}
\caption{For a TASEP with two sites, all non-vanishing $w_{j}^{i}$
are indicated by arrows in (a). (b) The only directed tree contributing to 
$P_{1}^{\ast}$. (c,d) The two directed trees contributing to 
$P_{3}^{\ast}$. }
\label{fig-5}
\end{figure}

The totally asymmetric exclusion process (TASEP ) on a one-dimensional chain
with open boundaries \cite{earlyTASEP,Schuetz} enjoys considerable attention
as a non-trivial system with a known NESS distribution. The dynamical rules
are as follows:\ Particles may enter (leave) a lattice with $L$ sites at the
left (right) end with rate $\alpha $ ($\beta $), provided the first (last)
site is empty (occupied). Within the system, particles may hop to the right,
with rate $\gamma $, provided the target site is empty. Though most of the
interesting properties appear in the thermodynamic limit, our goal here is
to illustrate the considerations of the previous sections. For that purpose,
it is more helpful to study a small system. To ensure that nontrivial loops
can exist in configuration space, the smallest `interesting' TASEP is
the one with two sites ($L=2$) and $4$ configurations. Let us label these as
follows: $1$ ($4$) for both sites being empty (full) and $2$ ($3$) for only
the right (left) site being occupied (Figure~\ref{fig-4}). 
In Figure~\ref{fig-5}(a), 
we show all the
allowed transitions (arrows). The rates are $\alpha $, 
$\beta $, and $\gamma $ for the vertical (blue), horizontal (green), 
and diagonal (black) arrows. The associated matrix $W$ rates
can be written easily:
\begin{equation}
W=\left( 
\begin{array}{cccc}
-\alpha  & \beta  & 0 & 0 \\ 
0 & -\left( \alpha +\beta \right)  & \gamma  & 0 \\ 
\alpha  & 0 & -\gamma  & \beta  \\ 
0 & \alpha  & 0 & -\beta 
\end{array}%
\right) \ .  \label{W-2tasep}
\end{equation}%
Note that it is ``maximally asymmetric'', in concordance with every
transition being uni-directional. As a result, most of the 16 possible trees
(cf.~Figure~\ref{fig-1}) 
do not contribute to the graph-theoretical representation of the stationary
probabilities $P_{i}^{\ast }$. For example, only a single tree remains 
in the representation of $P_{1}^{\ast}$ (shown in Figure~\ref{fig-5}(b)), and
only two trees (Figures \ref{fig-5}(c) and \ref{fig-5}(d)) remain for $P_{3}^{\ast}$
(see \ref{TASEP} for details).
The nonvanishing trees 
are easily found:\ $P_{1,2,3,4}^{\ast
}\propto \beta \gamma \beta ,\alpha \beta \gamma ,\alpha \alpha \beta +\beta
\beta \alpha ,\alpha \gamma \alpha $. For clarity, let us write the
normalized $P^{\ast }$ in dimensionless form:  
\begin{equation}
P^{\ast }=Z^{-1}\left( 
\begin{array}{c}
\beta /\alpha  \\ 
1 \\ 
\alpha /\gamma +\beta /\gamma  \\ 
\alpha /\beta 
\end{array}%
\right)   \label{P* 2tasep}
\end{equation}%
where $Z=1+$ $\alpha /\beta +\beta /\alpha +\alpha /\gamma +\beta /\gamma $
differs from $\mathcal{Z}$ in Equation (\ref{trees}) by a factor of $\alpha \beta
\gamma $. The stationary probability currents can be easily computed from
here. There are only two independent ones, e.g., $K^{\ast }{}_{4}^{2}=\alpha
/Z$ and $K^{\ast }{}_{1}^{2}=\beta /Z$. For these simple currents, there is
only one associated graph, each with an irreversible loop, illustrating 
Equation~(\ref{current-loops}). Again, the details are in \ref{TASEP}.

Turning to the dynamic equivalence classes associated with this process, the
simplest is to add an arbitrary $\Delta $ that satisfies Equation (\ref{Delta-db}%
), with 6 free parameters (denoted by $\epsilon _{1,...,6}$ here): 
\begin{equation*}
\fl
\Delta =  
\left( \begin{array}{cccc}
-\left( \epsilon _{1}+\epsilon _{2}+\epsilon _{3}\right) \frac{\alpha }{%
\beta } & \epsilon _{1} & \epsilon _{2}\frac{\gamma }{\alpha +\beta } & 
\epsilon _{3}\frac{\beta }{\alpha } \\ 
\epsilon _{1}\frac{\alpha }{\beta } & -\left( \epsilon _{1}+\epsilon
_{4}+\epsilon _{5}\right)  & \epsilon _{4}\frac{\gamma }{\alpha +\beta } & 
\epsilon _{5}\frac{\beta }{\alpha } \\ 
\epsilon _{2}\frac{\alpha }{\beta } & \epsilon _{4} & -\left( \epsilon
_{2}+\epsilon _{4}+\epsilon _{6}\right) \frac{\gamma }{\alpha +\beta } & 
\epsilon _{6}\frac{\beta }{\alpha } \\ 
\epsilon _{3}\frac{\alpha }{\beta } & \epsilon _{5} & \epsilon _{6}\frac{%
\gamma }{\alpha +\beta } & -\left( \epsilon _{3}+\epsilon _{5}+\epsilon
_{6}\right) \frac{\beta }{\alpha }%
\end{array} \right) 
\end{equation*}
Being the most general $\Delta $, the new transition matrix $W+\Delta $ does
not provide transparent insight. Nevertheless, we can draw some interesting
conclusions. For example, while the spectrum of $W$ is complex for some $%
\alpha $, $\beta $, and $\gamma $, the spectrum of $W+\Delta $ will be real
for sufficiently large $\epsilon $'s. Details for the case with only $%
\epsilon _{4}\neq 0$ are found in Appendix B. While this modification
corresponds to the rather innocuous addition of backwards hops for the
particle, more drastic changes can be made, such as $\epsilon _{3}$ which
allows particle pair creation/annihilation transitions! We reemphasize that,
in all cases, none of these modifications will affect the NESS distribution
of probabilities or their currents.

Finally, we illustrate the effect of $\Delta $ on entropy production. Since
the original $W$ does not allow ``backwards'' transitions, the production
rate is infinite. The addition of $\Delta $ changes this result to 
\begin{eqnarray}
\dot{\mathbf{S}}_{tot}^{\ast} = &Z^{-1} \left\{ \beta \ln \left( 1+\beta /\epsilon
_{1}\right) \left( 1+\beta /\epsilon _{2}\right) 
+\alpha \ln \left( 1+\alpha
/\epsilon _{5}\right) \left( 1+\alpha /\epsilon _{6}\right) \right. \nonumber \\
&+ \left. \left( \alpha
+\beta \right) \ln \left( 1+\left( \alpha +\beta \right) /\epsilon
_{4}\right) \right\} \,\,.
\end{eqnarray}%
$\allowbreak \,\,$Interestingly, this quantity is independent of $\epsilon
_{3}$, a result that follows from the lack of transitions between
configurations $1$ and $4$ in the original dynamics.

\subsection{Models of mass transport}
\label{examples-mass}

Motivated by a wide range of physical problems, simple models of mass
transport have been introduced. As extensions of TASEP, these models are
defined by more general rates for moving masses from site to site. Recently,
the exact steady-state distribution for a large class of such models was
found, so that they serve as good illustrations of the framework presented
above. Here, we will consider the most elementary model, the zero range
process (ZRP) \cite{ZRP}: discrete (but otherwise arbitrary) masses 
transported around a ring of discrete sites. Specifically, let each
site $s$ ($s=1,...,L$) of a periodic one-dimensional lattice be occupied by
an integer valued mass $m_{s}$ so that a configuration, $\mathcal{C}_{i}$ or
just $i$, is specified by the set $\left\{ m_{s}\right\} $. Thus, $%
w_{i}^{j}P_{j}^{\ast }$ in Equation (\ref{st-current}) would assume the form $%
P^{\ast }\left( \left\{ m_{s}\right\} \right) w\left( \left\{ m_{s}\right\}
\rightarrow \left\{ m_{s}^{\prime }\right\} \right) $. In the random
sequential updating version, a transport event takes place during an
infinitesimal time interval $dt$, at some site $k$: A portion, $\mu $, is
chipped off from $m_{k}$ and added to $m_{k+1}$, according to a given
conditional rate $q\left( \mu |m_{k}\right) $. In other words, the
transitions $w\left( \left\{ m_{s}\right\} \rightarrow \left\{ m_{s}^{\prime
}\right\} \right) $ connect only configurations with all $m$'s being
identical \emph{except} for a single pair. To be explicit, we have, 
\begin{equation}
m_{s}=m_{s}^{\prime }\quad {\rm for}\quad s\neq k,k+1  \label{mm'1}
\end{equation}%
while 
\begin{equation}
m_{k}-m_{k}^{\prime }=\mu ;\quad m_{k+1}-m_{k+1}^{\prime }=-\mu  \label{mm'2}
\end{equation}%
occurs with rate $q\left( \mu |m_{k}\right) $. This particular ``element''
in $P^{\ast }\left( \left\{ m_{s}\right\} \right) w\left( \left\{
m_{s}\right\} \rightarrow \left\{ m_{s}^{\prime }\right\} \right) $ is just 
\begin{equation*}
q\left( m_{k}-m_{k}^{\prime }|m_{k}\right) P^{\ast }\left( \left\{
m_{s}\right\} \right) \delta \left( m_{k}+m_{k+1},m_{k}^{\prime
}+m_{k+1}^{\prime }\right) \prod_{s\neq k,k+1}\delta \left(
m_{s},m_{s}^{\prime }\right) ,
\end{equation*}%
where $\delta \left( \bullet ,\bullet \right) $ is the Kronecker delta.
Clearly, the total mass $M\equiv \sum_{s}m_{s}$ is conserved and, with
finite $L$ as well, the configuration space is both discrete and finite.

In general, finding $P^{\ast }$ from a given $q$ can be quite difficult.
Fortunately, it can be found for a wide class of such models \cite{EMZ1}: If
(and only if) $q\left( \mu |m\right) $ can be cast in the form 
\begin{equation}
q\left( \mu |m\right) =g\left( \mu \right) f\left( m-\mu \right) /f\left(
m\right) ,
\end{equation}%
where $g,f$ are two arbitrary positive functions, then $P^{\ast }$ is
factorizable. Explicitly, $P^{\ast }\left( \left\{ m_{s}\right\} \right)
=Z^{-1}\prod_{s}f\left( m_{s}\right) $, where $Z\equiv \sum_{\left\{
m_{s}\right\} }\delta \left( M,\sum_{s}m_{s}\right) \prod_{s}f\left(
m_{s}\right) $ is a normalization factor. In configuration space 
(hypercubic $L$-dimensional lattice
$\left\{ m_{s}\right\} $),
the steady-state current 
\begin{equation}
\fl K^{\ast }\left( \left\{ m_{s}\right\} \rightarrow \left\{ m_{s}^{\prime
}\right\} \right) =P^{\ast }\left( \left\{ m_{s}\right\} \right) w\left(
\left\{ m_{s}\right\} \rightarrow \left\{ m_{s}^{\prime }\right\} \right)
-P^{\ast }\left( \left\{ m_{s}^{\prime }\right\} \right) w\left( \left\{
m_{s}^{\prime }\right\} \rightarrow \left\{ m_{s}\right\} \right)
\label{K*mm}
\end{equation}%
can exist only on planes spanned by $m_{k}$ and $m_{k+1}$, i.e., those
specified by Equations (\ref{mm'1}) and (\ref{mm'2}). The expressions reduce considerably,
since the $f$'s from $q$ and $P^{\ast }$ cancel.

If $m_{k}>m_{k}^{\prime }$, then the current is easily understood as the
result of moving $\mu >0$: 
\begin{eqnarray*}
K^{\ast }\left( m_{k},m_{k+1}\rightarrow m_{k}^{\prime },m_{k+1}^{\prime
}\right)  &=&Z^{-1} g\left( \mu \right) f\left( m_{k}-\mu \right) 
\prod_{s\neq k}f\left( m_{s}\right)  \\
&=&Z^{-1} g\left( m_{k}-m_{k}^{\prime }\right) f\left( m_{k}^{\prime }\right)
\prod_{s\neq k}f\left( m_{s}\right) 
\end{eqnarray*}%
Alternatively, we can keep $P^{\ast }$ implicit and write 
\begin{equation*}
K^{\ast }\left( m_{k},m_{k+1}\rightarrow m_{k}^{\prime },m_{k+1}^{\prime
}\right) =g\left( m_{k}-m_{k}^{\prime }\right) \frac{f\left( m_{k}^{\prime
}\right) }{f\left( m_{k}\right) }P^{\ast }\left( \left\{ m_{s}\right\}
\right)\,.
\end{equation*}%
In either case, a simple mnemonic for $K^{\ast }$ is: ``Replace $f\left(
m_{k}\right) $ by $g\left( m_{k}-m_{k}^{\prime }\right) f\left(
m_{k}^{\prime }\right) $ in $P^{\ast }$.'' For completeness, we should point
out that, for $m_{k}<m_{k}^{\prime }$, there is no transition across the $%
(k,k+1)$ bond, of course. However, the \emph{net current} is still
non-trivial, due to the \emph{difference} of terms 
in Equation (\ref{K*mm}). Thus, we
write 
\begin{eqnarray*}
K^{\ast }\left( m_{k},m_{k+1}\rightarrow m_{k}^{\prime },m_{k+1}^{\prime
}\right)  &=&- Z^{-1}g\left( m_{k}^{\prime }-m_{k}\right) f\left( m_{k}\right)
 \prod_{s\neq k}f\left( m_{s}^{\prime }\right)  \\
&=&-g\left( m_{k}^{\prime }-m_{k}\right) \frac{f\left( m_{k}\right) }{%
f\left( m_{k}^{\prime }\right) }P^{\ast }\left( \left\{ m_{s}^{\prime
}\right\} \right)\,.
\end{eqnarray*}%
A more general version of this system, highlighting other subtleties of mass
transport models, can be found in \ref{mass}.

Before ending this subsection, let us illustrate how Equation (\ref{Delta-db})
can be exploited here, to define other transition rates which will lead to
the same $\left\{ P^{\ast },K^{\ast }\right\} $. Since $P^{\ast }$ is
explicitly known, we can easily add further mass transports events without
altering the final NESS. For example, in addition to chipping $\mu $ from $%
m_{k}$ and moving it to $m_{k+1}$ as above, we may chip off amount of mass, $%
\tilde{\mu}$, from some other site, $\ell $, and move it to another site, $%
\ell ^{\prime }$, with an arbitrary rate $\tilde{w}$, \emph{provided} that
we also make the reverse move with rate $\tilde{w}f\left( m_{\ell }-\tilde{%
\mu}\right) f\left( m_{\ell ^{\prime }}+\tilde{\mu}\right) /f\left( m_{\ell
}\right) f\left( m_{\ell ^{\prime }}\right) $, according to Equation 
(\ref{Delta-db}). Of course, $\tilde{\mu}$ cannot be completely arbitrary, 
since it must lie within $[0,m_{\ell }]$. 
Further, \emph{many} masses can be chipped off at various sites
and moved\emph{\ simultaneously }(i.e., allowing moves away from planes in
the $L$-dimensional space), so that almost arbitrary configuration changes
can take place. The NESS, as we characterize it, will remain invariant
provided the reverse moves occur with a rate obeying Equation (\ref{Delta-db}).
It is clear that such ``long-ranged'' moves will be much more efficient in
Monte Carlo simulations of the true $P^{\ast }$. This is especially critical
for systems that display a condensate, since, on the one hand, $P^{\ast }$
is translationally invariant (or totally symmetric under permutations of $s$%
), while, on the other hand, typical configurations are associated with a
condensate located at a \emph{particular} site. With the original rates, the
tunneling times for the condensate to ``move'' from one site to another,
required to restore the invariance of the true $P^{\ast }$, are prohibitively
long \cite{LuckGodreche}. In contrast,
with arbitrarily long range jumps in configuration space, it is possible to
design transition rates that readily sample the symmetric $P^{\ast }$. In
this sense, such transitions are comparable to the cluster algorithms which
proved so useful in simulation studies of equilibrium systems \cite%
{Swendsen-Wang}.

\subsection{Ising models coupled to two thermal baths}
\label{examples-TT}

Since its inception 80 years ago, the static Ising model has attracted the
spotlight of equilibrium statistical physics. To simulate \emph{dynamic}
behavior when the system is coupled to a thermal reservoir, many researchers
proposed an assortment of transition rates that, not surprisingly, obey
detailed balance `with respect to' the famous equilibrium distribution. The
two most important classes are physically motivated: Glauber dynamics \cite%
{Glauber63} involving flips of individual spins and Kawasaki spin-exchange %
\cite{KK66} which conserves the total magnetization (or particle number, in
the lattice gas formulation). Given that this model is so well studied in
equilibrium, it offers itself naturally to investigations of \emph{non}%
-equilibrium steady states. Since the 80's, a large variety of rates that 
\emph{violate} detailed balance have been introduced, leading to an
extremely rich array of NESS-Ising models. The uniformly driven lattice gas %
\cite{KLS,SZrev}, based on Kawasaki dynamics, is a prime example. Another
class of such NESS is the two-temperature Ising model. Based on Glauber 
spin-flip dynamics, it is coupled to thermal baths at \emph{two} different
temperatures \cite{GM-TTIS,NL+me-TTIS,SZrev}. Needless to say, multiple
baths can also be considered, but we focus on just two, for simplicity. Even
with this restriction, the possible implementations of such couplings are
seemingly endless \cite{ZRRZ,SZrev}. For our purposes here, we consider a
one-dimensional chain with spins coupled alternately to the two baths. The
advantage of this model is that, in the steady state, \emph{all}
correlations are known \cite{SS2} and energy fluxes through the system have
been computed \cite{ZRRZ}. Indeed, the full time-dependence is also
accessible \cite{MM-TTIS}!

As before, we denote the sites of a periodic one-dimensional lattice by $s$ ($%
s=1,...,L$). Each site carries an Ising spin $\sigma _{s}$, so that a
configuration, $\mathcal{C}_{i}$ or just $i$, is specified by the set $%
\left\{ \sigma _{s}\right\} $ (or just $\left\{ \sigma \right\} $). Of
course, the energy of a configuration is given by the usual form: $%
-J\sum_{s}\sigma _{s}\sigma _{s+1}$. Choosing, for convenience, an even $L$,
we couple the spins at the even/odd sites to two different thermal
reservoirs, with temperatures 
\begin{equation}
T_{e}\equiv T_{even\,\,s}\neq T_{odd\,\,s}\,\equiv T_{o}\,.
\end{equation}%
Using specifically heat bath dynamics, the master equation for this system
can be written as 
\begin{equation}
\fl \partial _{t}P\left( \left\{ \sigma \right\} ,t\right) =\sum_{\left\{ \sigma
^{\prime }\right\} }\left[ w\left( \left\{ \sigma ^{\prime }\right\}
\rightarrow \left\{ \sigma \right\} \right) P\left( \left\{ \sigma ^{\prime
}\right\} ,t\right) -w\left( \left\{ \sigma \right\} \rightarrow \left\{
\sigma ^{\prime }\right\} \right) P\left( \left\{ \sigma \right\} ,t\right) %
\right]  \label{TTIS-ME}
\end{equation}%
where 
\begin{equation*}
w\left( \left\{ \sigma ^{\prime }\right\} \rightarrow \left\{ \sigma
\right\} \right) =w_{0}\sum_{s}\left\{ \frac{1}{2}\left[ 1+\gamma _{s}\sigma
_{s}\frac{\sigma _{s-1}^{\prime }+\sigma _{s+1}^{\prime }}{2}\right]
\prod_{j\neq s}\delta \left( \sigma _{j},\sigma _{j}^{\prime }\right)
\right\} ,
\end{equation*}%
and 
\begin{equation*}
\gamma _{s}\equiv \tanh \frac{2J}{k_{B}T_{s}},
\end{equation*}%
and $w_{0}$ just sets a scale. Note that this dynamics is quite special,
since the transition rates $w$ are independent of the state of the spin to
be flipped. Furthermore, with random sequential updates, only a single spin
is affected at a time. Thus, the probability currents exist only along the
``edges'' of an $2^{L}$-dimensional cube (the corners of which represent the
full configuration space). To proceed, it is useful to define a spin-flip
operator $\mathcal{F}_{s}$, acting on a spin configuration $\left\{ \sigma
\right\} :$%
\begin{equation*}
\mathcal{F}_{s}\left\{ \sigma \right\}\equiv \left\{ ...\sigma _{s-1},-\sigma
_{s},\sigma _{s+1},...\right\}
\end{equation*}%
which just flips the spin at site $s$. Now, we can write the net stationary
currents, e.g., 
\begin{eqnarray}
K^{\ast }\left( \left\{ \sigma \right\} \rightarrow \mathcal{F}_{s}\left\{
\sigma \right\} \right) &=&\frac{w_{0}}{2}\left[ 1-\gamma _{s}\sigma _{s}\frac{%
\sigma _{s-1}+\sigma _{s+1}}{2}\right] P^{\ast }\left( \left\{ \sigma
\right\} \right) \nonumber \\
&-& \frac{w_{0}}{2}\left[ 1+\gamma _{s}\sigma _{s}\frac{\sigma
_{s-1}+\sigma _{s+1}}{2}\right] P^{\ast }\left( \mathcal{F}_{s}\left\{
\sigma \right\} \right) .
\end{eqnarray}

In the limit $T_{even\,\,s}\rightarrow T_{odd\,\,s}$, all the $\gamma $'s
are the same and $P^{\ast }\left( \left\{ \sigma \right\} \right) \propto
\exp \left[ \left( J/k_{B}T\right) \sum_{s}\sigma _{s}\sigma _{s+1}\right] $%
, so that $K^{\ast }\equiv 0$. For the \emph{non}-equilibrium case, a simple
and compact form for $P^{\ast }\left( \left\{ \sigma \right\} \right) $ is
yet to be discovered. In particular, from the known correlations, we can
deduce that the ``effective Hamiltonian'' would involve long range
interactions \cite{SS1}. Probably the most convenient representation,
due to Hilhorst \cite{HH}, is 
\begin{equation}
\fl P^{\ast }\left( \left\{ \sigma \right\} \right) =\frac{1}{1+\lambda ^{L}}%
\sum_{\left\{ \tau \right\} }\left[ \prod_{s}\frac{1+\lambda \tau _{s}\tau
_{s+1}}{2}\right] \left[ \prod_{even\,\,s}\frac{1+h_{e}\tau _{s}\sigma _{s}}{%
2}\right] \left[ \prod_{odd\,\,s}\frac{1+h_{o}\tau _{s}\sigma _{s}}{2}\right]
.  \label{HH}
\end{equation}%
where 
\begin{equation*}
\lambda \equiv \tanh \left[ \frac{1}{2}\tanh ^{-1}\sqrt{\gamma _{e}\gamma
_{o}}\right] \quad {\rm and} \quad
h_{e,o} \equiv \sqrt{\left( \gamma _{e}+\gamma _{o}\right) /2\gamma _{o,e}}%
\,\,.
\end{equation*}%
Note that this representation requires performing a configuration-like sum
over an auxiliary set of spins $\left\{ \tau \right\} $, so that it
resembles the partition function for a one-dimensional Ising chain in an
inhomogeneous magnetic field. Thus, it is quite involved to find the
probability currents $K^{\ast }$. Not surprisingly, they are not zero in
general. Deferring details to elsewhere, let us provide some illustrations
here.

If the neighboring spins are opposite ($\sigma _{s-1}\neq \sigma _{s+1}$),
the transition rates, $w$, for flipping $\sigma _{s}$ are the same in both
directions. For an equilibrium system, such a local environment implies $%
P^{\ast }\left( \left\{ \sigma \right\} \right) =P^{\ast }\left( \mathcal{F}%
_{s}\left\{ \sigma \right\} \right) $, so that the steady-state current is
trivially zero. By contrast, we can show \cite{promises} that 
\begin{equation*}
P^{\ast }\left( \left\{ \sigma \right\} \right) \neq P^{\ast }\left( 
\mathcal{F}_{s}\left\{ \sigma \right\} \right)
\end{equation*}%
for, e.g., the configuration $\left\{ \sigma \right\}$ :
$\sigma _{s-1}=-1$ with \emph{all} other spins positive,
when $T_{even\,\,s}\neq T_{odd\,\,s}$. Thus, the net current $K^{\ast
}\left( \left\{ \sigma \right\} \rightarrow \mathcal{F}_{s}\left\{ \sigma
\right\} \right) $ is non-zero. Of course, it is proportional to the
difference of the two temperatures and vanishes in the equilibrium limit.
Its explicit form is rather complex without being especially illuminating,
and will not be displayed here \cite{promises}. Instead, let us present the 
\emph{sum }of these currents, over all spins \emph{not involved}, i.e., $%
\left\{ \bar{\sigma}\right\} \equiv \left\{ ...,\sigma _{s-2},\sigma
_{s+2},...\right\} $. For this, we define a reduced probability 
\begin{equation}
p\left( \sigma _{s-1},\sigma _{s},\sigma _{s+1}\right) \equiv \sum_{\left\{ 
\bar{\sigma}\right\} }P\left( \left\{ \sigma \right\} \right)
\end{equation}%
so that the sum 
\begin{equation*}
\mathcal{K}\left( \sigma _{s}\rightarrow -\sigma _{s}|\sigma _{s-1},\sigma
_{s+1}\right) \equiv \sum_{\left\{ \bar{\sigma}\right\} }K^{\ast }\left(
\left\{ \sigma \right\} \rightarrow \mathcal{F}_{s}\left\{ \sigma \right\}
\right)
\end{equation*}%
is simply 
\begin{equation*}
\fl \frac{w_{0}}{2}\left[ 1-\gamma _{s}\sigma _{s}\frac{\sigma _{s-1}+\sigma
_{s+1}}{2}\right] p\left( \sigma _{s-1},\sigma _{s},\sigma _{s+1}\right) -%
\frac{w_{0}}{2}\left[ 1+\gamma _{s}\sigma _{s}\frac{\sigma _{s-1}+\sigma
_{s+1}}{2}\right] p\left( \sigma _{s-1},-\sigma _{s},\sigma _{s+1}\right)
\end{equation*}%
From (\ref{HH}), it is easy to show that, in the $L\rightarrow \infty $
limit (where $\lambda ^{L}\rightarrow 0$), 
\begin{eqnarray*}
\fl p\left( \sigma _{s-1},\sigma _{s},\sigma _{s+1}\right) &\rightarrow &\frac{1%
}{2}\sum_{\tau _{s}}\left[ \frac{1+\lambda h_{s-1}\tau _{s}\sigma _{s-1}}{2}%
\right] \left[ \frac{1+h_{s}\tau _{s}\sigma _{s}}{2}\right] \left[ \frac{%
1+\lambda h_{s+1}\tau _{s}\sigma _{s+1}}{2}\right]  \nonumber \\
&=&\frac{1}{8}\left[ 1+\lambda h_{e}h_{o}\left( \sigma _{s-1}+\sigma
_{s+1}\right) \sigma _{s}+\left( \lambda h_{s-1}\right) ^{2}\sigma
_{s-1}\sigma _{s+1}\right]
\end{eqnarray*}%
where we have used $h_{s-1}=h_{s+1}$ and $h_{s-1}h_{s}=h_{e}h_{o}$. As a
result, 
\begin{equation*}
\fl \mathcal{K}\left( \sigma _{s}\rightarrow -\sigma _{s}|\sigma _{s-1},\sigma
_{s+1}\right) \rightarrow \frac{w_{0}}{8}\left[ 2\lambda h_{e}h_{o}-\gamma
_{s}\left( 1+\lambda ^{2}h_{s-1}^{2}\right) \right] \left\{ \sigma
_{s}\left( \sigma _{s-1}+\sigma _{s+1}\right) /2\right\} .
\end{equation*}%
Note that the last factor assumes the values $\pm 1,0$ only. After some
algebra, we find a particularly simple result: 
\begin{equation}
\left| \mathcal{K}\left( \sigma _{s}\rightarrow -\sigma _{s}|\sigma
_{s-1},\sigma _{s+1}\right) \right| =\left\{ 
\begin{array}{cc}
\left| \gamma _{e}-\gamma _{o}\right| \left( w_{0}/16\right) & {\rm for} %
\quad \sigma _{s-1}=\sigma _{s+1} \\ 
0 & {\rm for} \quad \sigma _{s-1}\neq \sigma _{s+1}%
\end{array}%
\right.
\end{equation}%
Note that this current \emph{sum} vanishes for $\sigma _{s-1}\neq \sigma
_{s+1}$, even though individual currents may be non-zero. Such a result is
not surprising, given that the sum restores an underlying up-down symmetry.
Finally, we remark that all quantities for finite $L$ are available, and this
expression turns out to be \emph{exact} for any $L$. Details of these
findings, as well as extensions such as computing current \emph{loops,} will
be published elsewhere \cite{promises}.

We end this subsection by illustrating how the currents can be used for
computing fluxes. In particular, we recover the results of \cite{ZRRZ}
concerning the energy flow into/out of the spin chain from the bath with the
higher/lower temperature. When one spin (say, $\sigma _{s}$) is flipped,
the energy change in the system is just 
\begin{equation*}
\Delta \mathcal{H}\left( \sigma _{s}\rightarrow -\sigma _{s}\right)
=2J\sigma _{s}\left( \sigma _{s-1}+\sigma _{s+1}\right) .
\end{equation*}%
Since this change is independent of all the other spins, we can simply
multiply it with $\mathcal{K}\left( \sigma _{s}\rightarrow -\sigma
_{s}\right) $ to obtain the rate of \emph{net }change (in the steady state).
The result is $w_{0}J\left( \gamma _{s-1}-\gamma _{s}\right) \left( \sigma
_{s-1}+\sigma _{s+1}\right) ^{2}/8$, so that the configurational average is 
$w_{0}J\left( \gamma _{s-1}-\gamma _{s}\right)$.
Note that, if $T_{s}>T_{s-1}$, then $\gamma _{s-1}>\gamma _{s}$ so that
there is an average energy flow ``into'' the system due to flipping the spin
at site $s$. Thus, the net flux \emph{through }our system is $w_{0}J\left|
\gamma _{s-1}-\gamma _{s}\right| /2$, a result phrased as ``the energy flow
from an even to an odd site'' in \cite{ZRRZ}. Interestingly, these energy
fluxes are directly related to the entropy production \cite{SS1,ZRRZ}.

\subsection{NESS Gaussian distributions}
\label{examples-gauss}

Another exactly solvable system is $\mathcal{N}$ ``particles'' (or simply,
degrees of freedom), subjected to linear forces (e.g., generalized coupled
harmonic oscillators) and Gaussian noise. Although the configuration space
of this system is continuous, it is sufficiently simple that both $P^{\ast }$
and $K^{\ast }$ are explicitly known, serving as an elegant illustration of
our formalism. It can be regarded as the continuum limit of a particular
biased random walk on, say, a hypercubic lattice. Specifically, the jump
rates have a bias which depends linearly on the location of the walker from
the origin \emph{and} may be anisotropic.

We begin with a Langevin equation for the degrees of freedom
(``coordinates'') $\xi ^{\mu }$ ($\mu =1,...,\mathcal{N}$) and the
associated Fokker-Planck equation for $P\left( \vec{\xi},t\right) $. The
latter is the master equation for this system, with a stationary
distribution known to be Gaussian. If detailed balance is violated, the
stationary probability currents can nevertheless be computed. 
This subsection will be
devoted to the highlights, with details left to \ref{gauss}. In this spirit,
we will use a compact notation of vectors (e.g., $\vec{\xi}$) and matrices
(blackboard bold font) here, while in \ref{gauss}, all indices will be
explicitly displayed.

Consider a Langevin equation with linear deterministic forces: 
\begin{equation}
\partial _{t}\vec{\xi}\left( t\right) =-\mathbb{F}\vec{\xi}\left( t\right) +%
\vec{\eta}\left( t\right)
\end{equation}%
and Gaussian noise (uncorrelated in time): 
\begin{eqnarray*}
\left\langle \vec{\eta}\left( t\right) \right\rangle &=&0 \\
\left\langle \vec{\eta}\left( t\right) \otimes \vec{\eta}\left( t^{\prime
}\right) \right\rangle &=&\mathbb{N}\delta \left( t-t^{\prime }\right) .
\end{eqnarray*}%
Here, $\mathbb{F}$ is an arbitrary real matrix, except that, for stability
and to model dissipation, the real part of its spectrum must be positive. We
denote the eigenvalues and the right and left eigenvectors of $F$ by 
\begin{equation}
\mathbb{F}\vec{u}_{I}=\lambda _{I}\vec{u}_{I}\quad {\rm and} 
\quad \vec{v}_{I}\mathbb{F}%
=\lambda _{I}\vec{v}_{I}
\end{equation}%
with $\mathop{\rm Re} \lambda _{I}>0$, and $I=1,2,...,\mathcal{N}$. 
Meanwhile, the
noise matrix $\mathbb{N}$ must clearly be real symmetric, as well as
positive. For example, if the discrete random walker has anisotropic jump
rates, then $\mathbb{N}$ is diagonal but not proportional to the unit matrix.

The associated Fokker-Planck equation reads 
\begin{equation}
\partial _{t}P\left( \vec{\xi},t\right) =\vec{\nabla}\cdot \left\{ \frac{%
\mathbb{N}}{2}\vec{\nabla}+\mathbb{F}\vec{\xi}\right\} P\left( \vec{\xi}%
,t\right)  \label{FP-G}
\end{equation}%
with the probability currents 
\begin{equation}
\vec{K}\equiv -\left\{ \frac{\mathbb{N}}{2}\vec{\nabla}+\mathbb{F}\vec{\xi}%
\right\} P\left( \vec{\xi},t\right) .
\end{equation}%

\emph{If} the matrices $\mathbb{N}$ and $\mathbb{F}$ are constrained such
that $\mathbb{N}^{-1}\mathbb{F}$ is \emph{symmetric}, then we can define 
\begin{equation}
\mathbb{V}\equiv 2k_{B}T\mathbb{N}^{-1}\mathbb{F}  \label{G-db}
\end{equation}%
and see that this is just a system of coupled simple harmonic oscillators,
subjected to (a general Gaussian) noise. Such a system will settle into
thermal equilibrium, with the familiar Boltzmann distribution: 
\begin{equation*}
P^{eq}\left( \vec{\xi}\right) \propto \exp \left\{ -\frac{\vec{\xi}\mathbb{V}%
\vec{\xi}}{2k_{B}T}\right\} 
\end{equation*}%
as well as a trivially vanishing $\vec{K}$. Of course, Equation (\ref{G-db}) with 
\emph{symmetric} $\mathbb{V}$ is just the detailed balance condition here
and expresses the usual fluctuation-dissipation relation.

However, if we insist on studying the general case where $\mathbb{N}^{-1}%
\mathbb{F}$ is \emph{not} symmetric, we will encounter a NESS with
non-trivial $\vec{K}$'s. (For $\mathcal{N}=3$, we would write such a $\vec{K}
$ as $curl$ of some vector field.) To see this explicitly, we recognize that
the stationary distribution is still a Gaussian \cite{ML} 
\begin{equation}
P^{\ast }\left( \vec{\xi}\right) \propto \exp \left[ -\frac{\vec{\xi}\mathbb{%
C}^{-1}\vec{\xi}}{2}\right] \,\,,  \label{eq:PG}
\end{equation}%
where $\mathbb{C}$ is the correlation matrix: 
\begin{equation}
\left\langle \vec{\xi}\otimes \vec{\xi}\right\rangle =\mathbb{C}.
\end{equation}%
To find $\mathbb{C}$ in terms of $\mathbb{N}$ and $\mathbb{F}$ is
straightforward, but not trivial (\ref{gauss}). A convenient expression is: 
\begin{equation}
\mathbb{C}=\sum_{I,J}\frac{\left\langle \vec{v}_{I}\mathbb{N}\vec{v}%
_{J}\right\rangle }{\lambda _{I}+\lambda _{J}}\vec{u}_{I}\otimes \vec{u}%
_{J}\,\,,  \label{C-answer}
\end{equation}%
which is manifestly symmetric.

Turning to the stationary currents, we see that the second term in 
\begin{equation}
\vec{K}^{\ast }\equiv -\left\{ \frac{\mathbb{N}}{2}\vec{\nabla}+\mathbb{F}%
\vec{\xi}\right\} P^{\ast }\left( \vec{\xi}\right)
\end{equation}%
can be written as $-\mathbb{FC}\vec{\nabla}P^{\ast }$.
Thus, $\vec{K}^{\ast }\equiv \left\{ \mathbb{FC}-\mathbb{N}/2\right\} \vec{%
\nabla}P^{\ast }$. To see that $\mathbb{FC}-\mathbb{N}/2$ is indeed \emph{%
antisymmetric} (so that $\vec{\nabla}\cdot \vec{K}^{\ast }=0$), we only need
to recall that $\vec{u}_{I}$ are eigenvectors of $\mathbb{F}$, so that 
\begin{equation*}
\mathbb{FC}=\sum_{I,J}\frac{\left\langle \vec{v}_{I}\mathbb{N}\vec{v}%
_{J}\right\rangle }{\lambda _{I}+\lambda _{J}}\lambda _{I}\vec{u}_{I}\otimes 
\vec{u}_{J}\,\,.
\end{equation*}%
The final result is quite simple: 
\begin{equation}
\vec{K}^{\ast }=\mathbb{A}\vec{\nabla}P^{\ast }\,\,,  \label{KAgP}
\end{equation}%
where 
\begin{equation}
\mathbb{A}\equiv \frac{1}{2}\sum_{I,J}\frac{\lambda _{I}-\lambda _{J}}{%
\lambda _{I}+\lambda _{J}}\left\langle \vec{v}_{I}\mathbb{N}\vec{v}%
_{J}\right\rangle \vec{u}_{I}\otimes \vec{u}_{J}  \label{A-answer}
\end{equation}%
is manifestly antisymmetric. Of course, we can evaluate the gradient in Equation (%
\ref{KAgP}) explicitly, with the result 
\begin{equation}
\vec{K}^{\ast }=-\left( \mathbb{AC}^{-1}\vec{\xi}\right) P^{\ast }\,\,.
\end{equation}%
One interesting consequence is that the expression in Equation (\ref{K^2}) can be
neatly evaluated: 
\begin{equation*}
\int \left| \vec{K}^{\ast }\right| ^{2}=\frac{1}{2}Tr\left\{ \mathbb{AC}^{-1}%
\mathbb{A}^{\top} \right\} \,\,.
\end{equation*}%

Lastly, let us consider ``generalized detailed balance'': What is the class
of dynamics that will lead us to a given NESS, i.e., $\left\{ P^{\ast
},K^{\ast }\right\} $ or more specifically in this case, $\left\{ \mathbb{C},%
\mathbb{A}\right\} $? Since the configuration space here is not finite, it
does not seem trivial to provide the \emph{entire} class of allowable
dynamics. However, since we know $P^{\ast }$ explicitly, we can exploit Equation (%
\ref{Delta-db}) and add $\int d\vec{\xi}^{\prime }\left\{ \Delta (\vec{\xi},%
\vec{\xi}^{\prime })P(\vec{\xi}^{\prime },t)-\Delta (\vec{\xi}^{\prime },%
\vec{\xi})P(\vec{\xi},t)\right\} $ to the right hand side of Equation (\ref{FP-G}%
), where $\Delta \left( \vec{\xi},\vec{\xi}^{\prime }\right) $ is \emph{any
function }that is non-negative and satisfies 
\begin{equation*}
\frac{\Delta \left( \vec{\xi},\vec{\xi}^{\prime }\right) }{\Delta \left( 
\vec{\xi}^{\prime },\vec{\xi}\right) }=\exp \left\{ \frac{\vec{\xi}^{\prime }%
\mathbb{C}^{-1}\vec{\xi}^{\prime }-\vec{\xi}\mathbb{C}^{-1}\vec{\xi}}{2}%
\right\} \,\,.
\end{equation*}

Unfortunately, this expression is too general to be illuminating. To provide
some insight, let us illustrate this freedom by studying a small subset of
the allowable modifications. Let us frame this question as the ``inverse''
of the usual one. Ordinarily, we are \emph{given} the matrices $\left\{ 
\mathbb{F},\mathbb{N}\right\} $, and are asked to find the NESS (in this
case, $\left\{ \mathbb{C},\mathbb{A}\right\} $). The inverse question is:
Given $\left\{ \mathbb{C},\mathbb{A}\right\} $, what can be said about $%
\left\{ \mathbb{F},\mathbb{N}\right\} $, i.e., what dynamics will insure
that we arrive at the above $\left\{ P^{\ast },K^{\ast }\right\} $? Clearly,
the key equation is just $\mathbb{FC-N}/2=\mathbb{A}$. 
Specifically, we are allowed to choose \emph{any} $\mathbb{N}$, provided
it is a valid noise correlation (real, symmetric, positive matrix) and
construct the dissipative forces $\mathbb{F}$ according to 
\begin{equation}
\mathbb{F}=\mathbb{C}^{-1}\left\{ \mathbb{N}/2+\mathbb{A}\right\} \,\,.
\label{FNCA}
\end{equation}
It may also appears as if we
can choose any matrix $\mathbb{F}$ and simply compute $\mathbb{N}$ from 
\begin{equation}
\mathbb{N}=2\left\{ \mathbb{FC-A}\right\} \,.  \label{NFCA}
\end{equation}
A closer examination reveals a difficulty with this naive approach. For
fixed $\left\{ \mathbb{C},\mathbb{A}\right\} $, not every choice of 
$\mathbb{F}$ will lead to a \emph{symmetric }$\mathbb{N}$ 
(a necessary condition for the noise correlation)! This issue is related to
the discussion of Equation (\ref{G-db}). 

We end this subsection with two comments. The concerned reader may ask: Why
can we choose an arbitrary $\mathbb{N}$ but not any $\mathbb{F}$? The
difference lies in that $\mathbb{N}$ has fewer parameters than $\mathbb{F}$.
There is an intimate connection to the general case where only the symmetric
part of $\bar{W}_{i}^{j}$ can be chosen freely. Since $\mathbb{N}$ must be
symmetric, it evidently contains all the freedom of choice available (in
this subclass of dynamics). By contrast, $\mathbb{F}$ appears to have no
particular symmetry; yet the antisymmetric part of $\mathbb{CF}$ is
completely fixed (by $\mathbb{A}$). Finally, to make contact with ordinary
detailed balance, it is easier to consider Equation (\ref{NFCA}) and follow how it reduces for the textbook case of an equilibrium system with the most
trivial dynamics: We have 
(i) a Hamiltonian $\mathcal{H}=\vec{\xi}\mathbb{V}\vec{\xi}/2$; 
(ii) forces $\mathbb{\kappa }\left( -\vec{\nabla}\mathcal{H}\right) $
where $\kappa $ is a dissipative coefficient; and 
(iii) $\mathbb{C}=\left( \mathbb{V}/k_{B}T\right) ^{-1}$. Thus, 
$\mathbb{A}\equiv 0$; and $\mathbb{N} = 2 \mathbb{FC}$ is just 
$2\kappa k_{B}T\,\times$ the unit matrix.

\section{ Summary and Outlook}

\label{summ}

In this paper, we propose that a \emph{non-equilibrium} steady state be
characterized not only by $P^{\ast }\left( \mathcal{C}\right) $, the
stationary distribution for finding the system in configuration $\mathcal{C}$%
, but also $K^{\ast }\left( \mathcal{C},\mathcal{C}^{\prime }\right) $, the 
\emph{net} probability current from $\mathcal{C}$ to $\mathcal{C}^{\prime }$%
. The pair $\left\{ P^{\ast },K^{\ast }\right\} $ forms a natural
generalization of having $P^{eq}$ as the quantity that characterizes all
details of an equilibrium state. Within the context of a master equation
approach, in which the dynamics of a system is defined by a set of
transition rates $w\left( \mathcal{C}\rightarrow \mathcal{C}^{\prime
}\right) $, we recalled a graphic method for constructing the pair $\left\{
P^{\ast },K^{\ast }\right\} $, as well as the intimate relationship between $%
K^{\ast }$ and irreversible cycles associated with the rates. Assuming
ergodic rates, this steady state is unique.

Exploring the consequences of our postulate, we find that the converse -
finding a set of $w$'s which leads to a given $\left\{ P^{*},K^{*}\right\} $
- is far from unique: The rates are only bound by a certain set of
constraints. This condition is well known, to those who study 
\emph{equilibrium} systems by computer simulations, 
as  ``detailed balance:'' 
\[
w\left( \mathcal{C}\rightarrow \mathcal{C}^{\prime }\right) P^{eq}\left( 
\mathcal{C}\right) -w\left( \mathcal{C}^{\prime }\rightarrow \mathcal{C}%
\right) P^{eq}\left( \mathcal{C}^{\prime }\right) =0
\]
The proposed generalization to \emph{non-}equilibrium systems is 
almost intuitive, namely, 
\[
w\left( \mathcal{C}\rightarrow \mathcal{C}^{\prime }\right) P^{*}\left( 
\mathcal{C}\right) -w\left( \mathcal{C}^{\prime }\rightarrow \mathcal{C}%
\right) P^{*}\left( \mathcal{C}^{\prime }\right) =K^{*}\left( \mathcal{C},%
\mathcal{C}^{\prime }\right) \,\,.
\]
Since the only difference between this and the equilibrium condition is the
value of the right hand side, we are ``just as free'' in choosing rates in
either case. In other words, the equivalence classes of rates are the same,
except for one subtlety explained in Section \ref{cons}. 
A unified way of making
this statement is the following. Two sets of rates belong to the same class
provided their \emph{differences,} $\Delta \left( \mathcal{C}\rightarrow 
\mathcal{C}^{\prime }\right) $, satisfy $\Delta \left( \mathcal{C}%
\rightarrow \mathcal{C}^{\prime }\right) P^{*}\left( \mathcal{C}\right)
=\Delta \left( \mathcal{C}^{\prime }\rightarrow \mathcal{C}\right)
P^{*}\left( \mathcal{C}^{\prime }\right) $. A further consequence of these
degrees of freedom is that the entropy production (associated with the
reservoirs coupled to our system, driving it to a NESS \cite{Schn}) can be
made infinitesimally small or arbitrarily large. We also proposed that $%
\Sigma \left( K^{*}\right) ^2$ can be exploited to measure of how ``far'' a
NESS is away from equilibrium. Of course, it would be better if a
\emph{dimensionless} quantity could be identified, 
so that common phrases like ``systems
far from equilibrium'' can be given a universal and quantitative meaning.
Finally, a number of examples were presented, illustrating these ideas.

While we hope that this work provides a fresh perspective on non-equilibrium
steady states, we are aware of much room for improvements and further
investigations. Let us end with a brief outlook.

For simplicity, we focused mainly on a master equation with continuous time
and discrete configuration space. Generalizations to systems with continuous
configuration space should be possible, though there may be non-trivial
mathematical obstacles. We believe that the example in \ref{examples-gauss}
represents a possible starting point. The more general case is the
Fokker-Planck equation, which admits transitions between configurations $%
\left\{ \phi \left( \vec{x}\right) \right\} $ that are
infinitesimally close: $\delta \phi \left( \vec{x}\right) $. Much work has
been devoted to this equation for NESS \cite{FP-general}, and the
resulting conclusions should be exploited. On a different note, formulations
involving discrete time present a different challenge. Since the transition
``rates'' are now conditional probabilities, their proper normalization will
lead to further constraints on the
choice of $w$'s, beyond those listed in Section \ref{cons}) . 
However, we anticipate this complication to be minor.

As more serious question, yet to be answered for arbitrary NESS's, is the
existence and uniqueness of thermodynamic limits. For example, there is a
belief that, for the driven Ising lattice gas in two dimensions, different
steady states will be reached in this limit, depending on the aspect ratio
of the system \cite{2dKLS}. Perhaps devoting some attention to the
distribution of probability currents will facilitate this quest. This topic
is also intimately related to the issue of the ``micro-macro connection.''
Starting as a vague notion of ``coarse graining,'' this connection has
gained much substance through the renormalization group (RG), especially in
the study of critical phenomena. For systems in equilibrium, the
distribution ($P^{eq}$) is uniquely linked to a Hamiltonian ($\mathcal{H}$%
) by the Boltzmann factor, and the progression from a microscopic, via a
mesoscopic, to a macroscopic description can be cast in the language of RG
flows in the space of $\mathcal{H}$'s or $P$'s. We believe that, for NESS,
the flow in the space of $K^{*}$'s will play an equally important role, on a
par with the flow of $P^{*}$. To illustrate this thought, let us highlight
two seemingly similar NESS systems in which the RG flows end at very
different types of fixed points. Specifically, for the uniformly driven
Ising lattice gas \cite{KLS}, the RG fixed point is a ``genuinely
non-equilibrium'' dynamic functional \cite{DDSfp} which \emph{cannot} be
written in the form of any equilibrium system with an $\mathcal{H}$. By
contrast, if the same system is driven randomly, the fixed point corresponds
to an equilibrium uniaxial system with dipolar interactions \cite{RDDSfp}.
This example suggests that the RG flow of $K^{*}$ leads to some non-trivial
fixed point in one case and, in the other case, ends at the ``trivial'' fixed
point $K^{*}=0$. One hint from physics is that there are global (particle)
currents in the first system, which cannot vanish under the RG. In contrast,
non-trivial microscopic probability currents in the second system are most
likely local, so that they vanish under RG transformations. Work is in
progress to provide solid foundations for these promising notions.

Entropy production is another venerable issue. Through its coupling to more
than one energy reservoir, our system can be regarded as an agent which
redistributes energy from one or more of these reservoirs to the rest. The
most intuitive and simple example is given in Section 
\ref{examples-TT} above - an
Ising model coupled to two thermals baths at different temperatures. Heat
flows from the hotter bath into the spin system and then to the cooler bath.
In the steady state, all thermodynamic quantities of the Ising system are
constant on the average, including this heat \emph{through-flux}. Thus,
there is a constant rate of energy redistribution associated with ``the
environment'' (or ``the medium''), a term we use to refer to the totality of 
\emph{all} the external reservoirs. Associated with this energy
redistribution should be an entropy production of the environment. In
general, given the detailed dynamics of the environment and how it couples
to our system, it is possible, in principle, to compute this rate of entropy
production. However, in our framework, we start with a master equation.
Expressed in terms of transition rates, it captures the dynamics of the
system and how the system couples to the environment, but carries no
information about the dynamics of the latter. As a result, it is not clear
how we can unambiguously find the rate of entropy production of the medium.
Nevertheless, it seems possible to define an entropy production, associated
with the system-medium \emph{coupling}. Relying on the context of chemical
reactions, Schnakenberg proposed such a definition, which depends only on
the $w$'s in the master equation. Thus, it is a natural candidate for us to
exploit here (see Section \ref{entropy}), even though its relation to the
true entropy production of the medium is not definitive. There may be other,
better grounded, definitions \cite{S-dot}. We believe it will be worthwhile
exploring them also in the context of $K^{*}$. In a related vein, we mention
fluctuation-dissipation and work-energy theorems associated with NESS \cite
{FT}. Apart from considerable theoretical interests over the last two
decades, recent experimental measurements in a variety of systems have
contributed to the excitement focused on these issues \cite{Expts}. 
Whether the probability currents provide novel insight here 
deserves to be studied.

Finally, let us return to the analogy with electrodynamics, in which a
parallel is drawn between electrostatics/magnetostatics and
equilibrium/non-equilibrium steady states. Of course, we are aware of
considerable conceptual difficulties, especially for the general case where
transitions between all configurations are allowed. However, for cases with
a local dynamics (e.g., Glauber spin flips for an Ising model or a
Fokker-Planck equation in continuous configuration space), an induced metric
can be defined \cite{RGraham} so that the probability current is just a
vector field. Since $K^{*}$ is divergence-free, it can be expressed as a
(generalized) curl of a vector potential and `magnetic fields' can also be
defined \cite{RGraham}. It behooves us to ask whether such fields might be
meaningful, and if their properties can be exploited. Pursuing this parallel
with electrodynamics far into the future, we may seek an underlying gauge
theory \cite{Timm} and see if it might provide a unified description of
all time-dependent phenomena in statistical mechanics.

The comments above highlight just a few of the issues surrounding the
central quest: providing a sound framework for the statistical mechanics of
non-equilibrium steady states, on a par with the Boltzmann-Gibbs approach
for equilibrium systems. We hope that our proposal - to bring the
microscopic probability current distribution, $K^{*}$, onto center stage, as
an equal partner with the microscopic distribution of probabilities, $P^{*}$
- will generate further explorations and discussions in the pursuit of our
quest.

\ack{} We thank U.~Seifert, H.~Spohn, J.~L.~Lebowitz, D.~Mukamel, 
M.~R.~Evans, R.~Harris, R.~Blythe, C. Timm, P.~Ao, and H.~Qian for
fruitful discussions and communications. Part of this work was performed
when the authors participated in a workshop at the Isaac Newton Institute,
Cambridge, UK. Financial support from the NSF through DMR-0414122 is 
gratefully acknowledged.

\appendix

\section {Kirchhoff's problem and its relation to NESS}
\label{Kirchhoff}

In most expositions of the graphic approach to finding the stationary
distribution from the master equation, Kirchhoff's solution \cite{Kirchhoff}
to the electrical network problem is referenced. Here, we offer some
comparisons and contrasts.

At first glance, it appears that both problems have the same underlying
network structure ($N$ vertices and $N\left( N-1\right) /2$ edges) and the
same number of parameters: $2\times N\left( N-1\right) /2$. For electrical
networks, Kirchhoff \cite{Kirchhoff} posed a resistance ($R_{ij}$) and an
electromotive force ($\mathcal{E}_{ij}$) between the nodes $i$ and $j$. The
former/latter are strictly symmetric/antisymmetric. Naively, since we have $%
w_{i}^{j}$ (with $i\neq j$), we may think of identifying the symmetric and
antisymmetric parts of $w$ with $R$ and $\mathcal{E}$, and perhaps formulate
an exact mapping between the two problems.

On closer examination, differences emerge. To seek the NESS, we need to
solve $N$ equations for $N$ unknowns ($P_{i}^{\ast }$). No other equations
need be solved to obtain the $N\left( N-1\right) /2$ currents; they are just
linear combinations of the $P^{\ast }$'s. By contrast, by focusing on the
currents ($I_{ij}$, also antisymmetric) as the unknowns, Kirchhoff needed
far more equations. In his paper, there are $N$ equations for the nodes: 
\begin{equation*}
\sum_{j}I_{ij}=0
\end{equation*}
for each $i$,
and $N\left( N-3\right) /2+1$ equations for the loops: 
\begin{equation*}
\mathcal{E}_{k_{1}k_{2}}+...+\mathcal{E}%
_{k_{n}k_{1}}-R_{k_{1}k_{2}}I_{k_{1}k_{2}}...-R_{k_{n}k_{1}}I_{k_{n}k_{1}}=0%
\,\,.
\end{equation*}%
Of course, we recognize that the node equations correspond to the master
equation for the steady state and that loop equations allow us to define a
single-valued function associated with each node. Traditionally, this is
called the ``potential,'' $V_{i}$, which can be defined through the
differences 
\begin{equation*}
V_{i}-V_{j}\equiv \mathcal{E}_{ij}-R_{ij}I_{ij}\,\,.
\end{equation*}%
Clearly this is not the only possible definition of the single-valued
function, since any well-behaved function of $V_{i}$ will produce another
single-valued function. Even if we demand linearity between $V$ and $I$, the
above is unique only up to two parameters: a constant shift and an overall
scale: $V\rightarrow \lambda \left( V-v\right) $. If we were to try to
identify $P_{i}^{\ast }$ with the potential, we will need the shift to
insure positivity (since potentials can easily be negative) and the scale
for normalization of $P^{\ast }$. But the shift is arbitrary (apart from a
lower bound), so that the ratios $\left( V_{i}-v\right) /\left(
V_{j}-v\right) $ are not fixed. In this sense, there is no way to generate a
unique quantity like $P_{i}^{\ast }$ from the currents of the Kirchhoff
problem.

Forcing the issue from the other direction seems more feasible, since there
is no problem choosing the potentials $V_i$ to be $P_i^{*}$. From the
definition $K^{*}{}_i^j\equiv w_i^jP_j^{*}-w_j^iP_i^{*}$, we can write 
\begin{equation*}
P_i^{*}=\frac{w_i^j}{w_j^i}P_j^{*}+\frac 1{w_j^i}K^{*}{}_i^j
\end{equation*}
Formally, we can substitute the expression in Equation (\ref{trees}) for $P_j^{*}$
and write the first term 
( $ w_i^j P_j^{*} / w_j^i $ )
as $G_i^j\left[ \left\{ w\right\} \right] $. Then,
we have 
\begin{equation*}
P_i^{*}-P_j^{*}=\left[ G_i^j-G_j^i\right] -\left[ \frac 1{w_j^i}+\frac
1{w_i^j}\right] K^{*}{}_j^i\,\,.
\end{equation*}
Taken at face value, we can exploit this equation to identify 
\begin{eqnarray*}
P_i^{*} &\rightarrow &V_i \\
K^{*}{}_j^i &\rightarrow &I_{ij} \\
G_i^j-G_j^i &\rightarrow &\mathcal{E}_{ij} \\
\frac 1{w_j^i}+\frac 1{w_i^j} &\rightarrow &R_{ij}
\end{eqnarray*}

The lack of a one-to-one mapping between $\left\{ R,\mathcal{E}\right\} $
and $\left\{ w\right\} $ can now be summarized succinctly. From the former,
we can find a unique solution for the currents, but not the charges (or
potentials). Even if we could and these were identified with $K^{\ast }$ and 
$P^{\ast }$, we know that there is a whole class of $w$'s which leads to the
very same $\left\{ P^{\ast },K^{\ast }\right\} $. Finally, in passing, we
mention that there are intimate connections between Kirchhoff's problem and
the percolation problem. Details may be found in the context of an extensive
review on the Potts model \cite{FYWu}.

\section{A maximally asymmetric $N=4$ example: TASEP with two
sites}
\label{TASEP}

For a totally asymmetric exclusion process (TASEP ) on a one-dimensional
chain with open boundaries \cite{earlyTASEP, Schuetz} with two sites, we show
all the configurations in Figure~\ref{fig-4}. 
From the transition matrix, Equation (\ref{W-2tasep}), 
we see that there are only 5 non-zero arrows in the full space,
shown in Figure \ref{fig-5}(a). 
As a result, there are very few non-trivial directed trees. In
Figure \ref{fig-5}(b), we show the only tree directed towards vertex $1$ that is
non-zero. For vertex $3$, there are two trees, shown in Figure \ref{fig-5}(c) and (d). 
Thus,
we arrive at, respectively, $P_{1,2,3,4}^{\ast }\propto \beta \gamma \beta
,\alpha \beta \gamma ,\alpha \alpha \beta +\beta \beta \alpha ,\alpha \gamma
\alpha $ , and $\mathcal{Z}$ in Equation (\ref{trees}) is just $\alpha ^{2}\beta
+\beta ^{2}\alpha +\left( \alpha ^{2}+\alpha \beta +\beta ^{2}\right) \gamma 
$. To illustrate Equation (\ref{current-loops}), we consider, e.g., 
$K^{\ast}{}_{3}^{1}$ which is $\beta /Z$. 
In graphic terms, we add a down arrow from
vertex $1$ to $3$ to Figure \ref{fig-5}(b), 
forming an irreversible loop associated with 
$\Pi \left[ \mathcal{L}\right] =\alpha \beta \gamma $. Meanwhile, there is
only one side branch, so that $R=\beta $ here. Thus, $K^{\ast
}{}_{3}^{1}=\Pi R/\mathcal{Z}=\alpha \beta \gamma \beta /(\alpha \beta
\gamma Z)$, in agreement with the above.

From these $K^{\ast }$'s, we can evaluate the particle current flowing into
the system. Referring to Equation (\ref{Jbond}), it is clear that we just need 
to consider $\left( 1-n_{1}\right) n_{1}^{\prime }$,
which is zero for all configuration pairs, \emph{except} being unity for the 
$\left( 1,3\right) $ and $\left( 2,4\right) $ pairs. Thus,  
\begin{equation*}
\left\langle J^{\ast }\right\rangle =K^{\ast }{}_{3}^{1}+K^{\ast
}{}_{4}^{2}=\left( \alpha +\beta \right) /Z\ .
\end{equation*}%
In steady state, it is clear that this expression will also equal the
current flowing between the two sites: $K^{\ast }{}_{2}^{3}$, or out of the
system: $K^{\ast }{}_{1}^{2}+K^{\ast }{}_{3}^{4}$.

Next, we turn to the dynamic equivalence classes associated with this $W$.
We first compute the matrix $\bar{W}$, and then extract its symmetric and
its antisymmetric part, resulting in:\ $\allowbreak $ $\allowbreak $%
\begin{eqnarray}
S& =\frac{1}{2}Z^{-1}\left( 
\begin{array}{cccc}
-2\beta & \beta & \beta & 0 \\ 
\beta & -2\left( \alpha +\beta \right) & \alpha +\beta & \alpha \\ 
\beta & \alpha +\beta & -2\left( \alpha +\beta \right) & \alpha \\ 
0 & \alpha & \alpha & -2\alpha%
\end{array}%
\right) \label{S-TASEP}
\\
A& =\frac{1}{2}Z^{-1}\left( 
\begin{array}{cccc}
0 & \beta & -\beta & 0 \\ 
-\beta & 0 & \alpha +\beta & -\alpha \\ 
\beta & -\left( \alpha +\beta \right) & 0 & \alpha \\ 
0 & \alpha & -\alpha & 0%
\end{array}%
\right) \label{A-TASEP}
\end{eqnarray}%
We clearly recognize the values of the $K^{\ast }$'s 
in the elements of $A$.

Since $S$ is a $4\times 4$ symmetric matrix with constrained diagonal
elements, we have $6$ degrees of freedom. Selecting just one of these, for
the purposes of illustration, we can define a new matrix, $\hat{S}$, by,
e.g., modifying the $(3,2)$ and $(2,3)$ elements by an amount $2\epsilon $,
and preserving the column sums 
\begin{equation*}
\hat{S}=\frac{1}{2}Z^{-1}\left( 
\begin{array}{cccc}
-2\beta & \beta & \beta & 0 \\ 
\beta & -2\left( \alpha +\beta \right) -2\epsilon & \alpha +\beta +2\epsilon
& \alpha \\ 
\beta & \alpha +\beta +2\epsilon & -2\left( \alpha +\beta \right) -2\epsilon
& \alpha \\ 
0 & \alpha & \alpha & -2\alpha%
\end{array}%
\right)
\end{equation*}%
Respecting the symmetries and the constraint (\ref{M-const}), the new rate
matrix $\hat{W}$ is given by 
\begin{equation*}
\fl \hat{W}=\left[ \hat{S}_{i}^{j}+\frac{1}{2}K^{\ast }{}_{i}^{j}\right]
(P_{j}^{\ast })^{-1}=\left( 
\begin{array}{cccc}
-\alpha & \beta & 0 & 0 \\ 
0 & -\left( \alpha +\beta \right) -\epsilon & \gamma +\frac{\gamma }{\alpha
+\beta }\epsilon & 0 \\ 
\alpha & \epsilon & -(\alpha +\beta )-\frac{\gamma }{\alpha +\beta }\epsilon
& \beta \\ 
0 & \alpha & 0 & -\beta%
\end{array}%
\right)
\end{equation*}%
Comparing to the original matrix $W$, we can see that the new feature is a
transition from configuration $2$ to $3$ which was absent in the original
model. In other words, the particle is now allowed to jump backwards. The
matrix $\Delta $ is nonzero only in the $2\times 2$ submatrix involving
configurations $2$ and $3$: 
\begin{equation*}
\Delta \equiv \hat{W}-W=\epsilon \left( 
\begin{array}{cccc}
0 & 0 & 0 & 0 \\ 
0 & -1 & \frac{\gamma }{\alpha +\beta } & 0 \\ 
0 & 1 & -\frac{\gamma }{\alpha +\beta } & 0 \\ 
0 & 0 & 0 & 0%
\end{array}%
\right)
\end{equation*}%
It is easy to check that it satisfies Equation (\ref{Delta-db}).

Let us now recalculate the particle current from the first to the second
site, keeping in mind that the particle can now also jump backwards.
However, the probability current between configurations $2$ and $3$ is
unchanged, as one can see from a quick explicit check:
\begin{equation*}
K^{\ast }{}_{2}^{3}=\hat{w}_{2}^{3}P_{3}^{\ast }-\hat{w}_{3}^{2}P_{2}^{\ast
}=Z^{-1}\left[ \left( \gamma +\frac{\gamma }{\alpha +\beta }%
\epsilon \right) \frac{\left( \alpha +\beta \right) }{\gamma }-\epsilon %
\right] =Z^{-1}\left( \alpha +\beta \right)
\end{equation*}%
Since each term in the equation above is associated with a single particle
hop, it is immediately obvious that $\left\langle J^{\ast }\right\rangle $
also remains invariant, under the transformation from $W$ to $\hat{W}$. \ 

Allowing for backwards hops of the particle may appear rather innocuous. A
more drastic modification would be to allow transitions between
configurations $1$ and $4$ which are completely absent in the original
model. This leads to 
\begin{equation*}
\hat{S}=\frac{1}{2}Z^{-1}\left( 
\begin{array}{cccc}
-2(\beta+\epsilon) & \beta & \beta & 2\epsilon \\ 
\beta & -2\left( \alpha+\beta\right) & \alpha+\beta & \alpha \\ 
\beta & \alpha+\beta & -2\left( \alpha+\beta\right) & \alpha \\ 
2\epsilon & \alpha & \alpha & -2(\alpha+\epsilon)%
\end{array}
\right)
\end{equation*}
and 
\begin{equation*}
\hat{W}=W+\epsilon\left( 
\begin{array}{cccc}
-\alpha/\beta & 0 & 0 & \beta/\alpha \\ 
0 & 0 & 0 & 0 \\ 
0 & 0 & 0 & 0 \\ 
\alpha/\beta & 0 & 0 & -\beta/\alpha%
\end{array}
\right)
\end{equation*}
Again, the difference between the two sets of rates satisfies detailed
balance with respect to $P^{\ast}$. As a result, the probability current $%
K^{\ast}{}_{4}^{1}$, associated with transitions between configurations $1$
and $4$, remains zero. This also ensures that the physical current, $%
\left\langle J^{\ast}\right\rangle $, remains unaffected.

Since all of these rates involve at least one uni-directional transition,
they are all associated with infinite entropy production. In order to obtain
a finite value for the entropy production, we need to start from a rate
matrix where each forward transition is associated with a backward one.
Starting from the original $W$ and its associated $S$, Equation (\ref{S-TASEP}),
and recalling that we
have $6$ degrees of freedom, parameterized by $\epsilon _{1},$ $\epsilon
_{2},...,\epsilon _{6}$, the most general $\hat{S}$ would be 
\begin{eqnarray*}
\fl \hat{S}=S + Z^{-1}
\left( 
\begin{array}{cccc}
-\left( \epsilon _{1}+\epsilon _{2}+\epsilon _{3}\right) & 
\epsilon _{1} & \epsilon _{2} & \epsilon _{3} \\ 
\epsilon _{1} & -\left( \epsilon
_{1}+\epsilon _{4}+\epsilon _{5}\right) & \epsilon _{4} & 
\epsilon _{5} \\ 
\epsilon _{2} & \epsilon _{4} & 
-\left( \epsilon _{2}+\epsilon _{4}+\epsilon _{6}\right) & 
\epsilon _{6} \\ 
\epsilon _{3} & \epsilon _{5} & \epsilon _{6} & 
-\left( \epsilon _{3}+\epsilon _{5}+\epsilon _{6}\right)%
\end{array}%
\right)
\end{eqnarray*}%
The associated $\hat{W}$ is given by 
\begin{equation*}
\fl \hat{W}=W+ \left(%
\begin{array}{cccc}
-\left( \epsilon _{1}+\epsilon _{2}+\epsilon _{3}\right) \frac{\alpha }{%
\beta } & \epsilon _{1} & \epsilon _{2}\frac{\gamma }{\alpha +\beta } & 
\epsilon _{3}\frac{\beta }{\alpha } \\ 
\epsilon _{1}\frac{\alpha }{\beta } & -\left( \epsilon _{1}+\epsilon
_{4}+\epsilon _{5}\right) & \epsilon _{4}\frac{\gamma }{\alpha +\beta } & 
\epsilon _{5}\frac{\beta }{\alpha } \\ 
\epsilon _{2}\frac{\alpha }{\beta } & \epsilon _{4} & -\left( \epsilon
_{2}+\epsilon _{4}+\epsilon _{6}\right) \frac{\gamma }{\alpha +\beta } & 
\epsilon _{6}\frac{\beta }{\alpha } \\ 
\epsilon _{3}\frac{\alpha }{\beta } & \epsilon _{5} & \epsilon _{6}\frac{%
\gamma }{\alpha +\beta } & -\left( \epsilon _{3}+\epsilon _{5}+\epsilon
_{6}\right) \frac{\beta }{\alpha }%
\end{array} \right)%
\end{equation*}%
Evaluating the entropy production for this general form, we arrive at 
\begin{eqnarray*}
\dot{\mathbf{S}}_{tot}^{\ast }& =\frac{1}{2}\sum_{i,j}K^{\ast }{}_{i}^{j}\ln 
\frac{W_{i}^{j}P_{j}^{\ast }}{W_{j}^{i}P_{i}^{\ast }}=\frac{1}{2}%
\sum_{i,j}K^{\ast }{}_{i}^{j}\ln \frac{S_{i}^{j}+A_{i}^{j}}{%
S_{i}^{j}-A_{i}^{j}} \\
 &=Z^{-1} \left\{ \beta \ln \frac{\beta +\epsilon _{1}}{\epsilon _{1}}+\beta \ln \frac{%
\beta +\epsilon _{2}}{\epsilon _{2}}+\left( \alpha +\beta \right) \ln \frac{%
\alpha +\beta +\epsilon _{4}}{\epsilon _{4}} \right. \\
& \quad \left. + \alpha \ln \frac{\alpha +\epsilon _{5}}{\epsilon _{5}}+\alpha \ln \frac{%
\alpha +\epsilon _{6}}{\epsilon _{6}} \right\}
\end{eqnarray*}%
which is clearly finite now.

\section{Currents in generalized mass transport models}
\label{mass}

Here, we provide a further example associated with mass transport models,
namely, the generalization from discrete to continuous masses, with parallel
updates in discrete time. Thus, we consider a ring of discrete sites
occupied by \emph{continuous} $m_{s}$'s. The evolution proceeds in discrete
time steps, with $\mu _{s}$ being chipped off from $m_{s}$ and added to $%
m_{s+1}$, for all sites at once. To distinguish the conditional probability
here from the case in the main text, we denote it by $\phi \left( \mu
|m\right) $ here. Though the fundamental issues for this non-equilibrium
process and the one with continuous time and random sequential updates are
the same, there are a few differences, on which we will comment here.

As in the previous case, $P^{\ast }$ is known for a wide class of such models%
\cite{EMZ1}: If $\phi \left( \mu |m\right) $ is of the form $\left. v\left(
\mu \right) u\left( m-\mu \right) \right/ f\left( m\right) $ (where $v$ and $%
u$ are arbitrary non-negative functions, and $f\left( m\right) =\int v\left(
\mu \right) u\left( m-\mu \right) d\mu $), then $P^{\ast }\left( \left\{
m_{s}\right\} \right) = Z^{-1} \prod_{s}f\left( m_{s}\right) $, as in Section
\ref{examples-mass}. Again, 
$w_{i}^{j}P_{j}^{\ast }$ in Equation (\ref{st-current}) will assume the form 
$P^{\ast }\left( \left\{ m_{s}\right\} \right) $ $w\left( \left\{
m_{s}\right\} \rightarrow \left\{ m_{s}^{\prime }\right\} \right) $. Unlike
the random sequential case, 
\begin{equation*}
w\left( \left\{ m_{s}\right\} \rightarrow \left\{ m_{s}^{\prime }\right\}
\right) =\prod_{s}\phi \left( \mu _{s}|m_{s}\right)
\end{equation*}%
so that \emph{all} the $f$'s cancel. Thus, we are left with a simple
product: 
\begin{equation*}
P^{\ast }\left( \left\{ m_{s}\right\} \right) w\left( \left\{ m_{s}\right\}
\rightarrow \left\{ m_{s}^{\prime }\right\} \right) =Z^{-1}\prod_{s}v\left(
\mu _{s}\right) u\left( \sigma _{s}\right)
\end{equation*}%
where 
$\mu _{s}$ (the mass to be moved) and $\sigma _{s}$ (the remaining mass) are
fixed by 
\begin{eqnarray}
m_{s}=\mu _{s}+\sigma _{s} \nonumber \\
m_{s}^{\prime }=\mu_{s-1}+\sigma _{s}  \label{mm'}
\end{eqnarray}%
Thus, 
\begin{equation}
ZK^{\ast }\left( \left\{ m_{s}\right\} \rightarrow \left\{ m_{s}^{\prime
}\right\} \right) =\prod_{s}v\left( \mu _{s}\right) u\left( \sigma
_{s}\right) -\prod_{s}v\left( \tilde{\mu}_{s}\right) u\left( \tilde{\sigma}%
_{s}\right) ,  \label{K*-mtm}
\end{equation}%
where $\tilde{\mu}_{s}$ and $\tilde{\sigma}_{s}$ satisfy $m_{s}^{\prime }=%
\tilde{\mu}_{s}+\tilde{\sigma}_{s}$ and $m_{s}=\tilde{\mu}_{s-1}+\tilde{%
\sigma}_{s}$. Let us end with a few remarks.

First, it is clear that many different functions $v$ and $u$ will lead to
the same convolution $f\equiv v\ast u$, so that all of them are associated
with the same $P^{\ast }$. However, it is also clear that typically, $%
K^{\ast }$ will \emph{not} be the same. Thus, our framework provides a
natural way to distinguish these different NESS's, all of which share the
same stationary $P^{\ast }$.

Next, we should emphasize that this rather compact notation belies a
subtlety: Starting with a pair of configurations $\left\{
m_{s},m_{s}^{\prime }\right\} $, it is not possible in general to find a
unique set $\left\{ \mu _{s},\sigma _{s}\right\} $. First, Equation (\ref{mm'}),
rewritten here as 
\begin{eqnarray*}
\mu _{s-1}+\sigma _{s} &=&m_{s}^{\prime } \\
\mu _{s}+\sigma _{s} &=&m_{s}
\end{eqnarray*}%
is just a simple linear equation of the form 
\begin{equation*}
\left( 
\begin{array}{ccccc}
1 & 0 & ... & 0 & 1 \\ 
1 & 1 & ... & 0 & 0 \\ 
\vdots & \vdots & \vdots & \vdots & \vdots \\ 
0 & 0 & ... & 1 & 0 \\ 
0 & 0 & ... & 1 & 1%
\end{array}%
\right) \left( 
\begin{array}{c}
\sigma _{1} \\ 
\mu _{1} \\ 
\vdots \\ 
\sigma _{L} \\ 
\mu _{L}%
\end{array}%
\right) =\left( 
\begin{array}{c}
m_{1}^{\prime } \\ 
m_{1} \\ 
\vdots \\ 
m_{L}^{\prime } \\ 
m_{L}%
\end{array}%
\right)
\end{equation*}%
Now, the matrix on the left has a zero eigenvalue and cannot, strictly, be
inverted. However, the zero eigenvector ($+1/-1$ for even/odd elements) is
associated with mass conservation. A physical set of $\left\{
m_{s},m_{s}^{\prime }\right\} $ obeys $\sum_{s}(m_{s}-m_{s}^{\prime })=0$
and has no component along this eigenvector. Thus, the existence of a
solution is not in doubt.

In general, the solution is not unique, however. With parallel update, the
transported masses $\left\{ \mu _{s},\sigma _{s}\right\} $ are determined by
configurational masses $\left\{ m_{s},m_{s}^{\prime }\right\} $ only up to
an overall parameter $\bar{\mu}$. This extra ``freedom'' is associated with
a simple physical process: The same configurational change can be achieved
if we further chip off an identical amount, $\bar{\mu}$, from \emph{all}
sites and moved to the next, i.e., replacing $\left\{ \mu _{s},\sigma
_{s}\right\} $ by $\left\{ \mu _{s}+\bar{\mu},\sigma _{s}-\bar{\mu}\right\} $%
. The range of allowed $\bar{\mu}$ is clear: None of the resultant $\mu $'s
and $\sigma $'s can be negative. Thus, if we start with any ``minimal'' set
of $\mu _{s}$'s (i.e., at least one $\mu $ is zero), then the other sets can
be generated by adding a positive $\bar{\mu}$, \emph{provided} it is no
larger than the smallest $m_{s}$ in the system. In other words, from any
valid solution (i.e., non-negative $\left\{ \mu _{s},\sigma _{s}\right\} $)
of Equation (\ref{mm'}), $\bar{\mu}$ can decreased (or increased) until the
smallest $\mu _{s}$ (or $\sigma _{s}$) vanishes. In this sense, we should
integrate over this allowed range of $\bar{\mu}$ when computing the first
term in the current in Equation (\ref{K*-mtm}).

Once a $\left\{ \mu _{s},\sigma _{s}\right\} $ solution is found, there is a
simple way to construct the ``complementary'' set $\left\{ \tilde{\mu}_{s},%
\tilde{\sigma}_{s}\right\} $, which satisfies 
\begin{eqnarray*}
\tilde{\mu}_{s-1}+\tilde{\sigma}_{s} &=&m_{s} \\
\tilde{\mu}_{s}+\tilde{\sigma}_{s} &=&m_{s}^{\prime }
\end{eqnarray*}%
Indeed, we can access the minimal set immediately. Let $\hat{\mu}$ be the
largest of the $\mu $'s: 
\begin{equation*}
\hat{\mu}\equiv \sup \left\{ \mu _{s}\right\} .
\end{equation*}%
Then, 
\begin{eqnarray*}
\tilde{\mu}_{s} &=&\hat{\mu}-\mu _{s} \\
\tilde{\sigma}_{s} &=&\sigma _{s}+\mu _{s}+\mu _{s-1}-\hat{\mu}.
\end{eqnarray*}%
Note that the one parameter family of allowed $\left\{ \tilde{\mu}_{s},%
\tilde{\sigma}_{s}\right\} $ is not necessarily the same as the one for $%
\left\{ \mu _{s},\sigma _{s}\right\} $. As a result, a different integral is
needed here to compute the second term of $K^{\ast }$ in Equation (\ref{K*-mtm}).

\section{A simple example with continuous configuration space.}
\label{gauss}

In the following, we provide the detailed analysis of the Langevin equation
with linear deterministic forces (generalization of a system of coupled
harmonic oscillators). The configuration space is $\mathcal{N}$-dimensional
space: $\left\{ \xi ^{\mu }\right\} $ with $\mu =1,...,\mathcal{N}$. The
Langevin equation for $\xi ^{\mu }\left( t\right) $ reads: 
\begin{equation*}
\partial _{t}\xi ^{\mu }=-F_{\nu }^{\mu }\xi ^{\nu }+\eta ^{\mu }
\end{equation*}%
where $F_{\nu }^{\mu }$ is an arbitrary real matrix, except that, for
stability, the real part its spectrum must be positive (dissipative forces).
The noise $\eta $ is Gaussian with zero mean, i.e., 
\begin{eqnarray*}
\left\langle \eta ^{\mu }\left( t\right) \right\rangle  &=&0 \\
\left\langle \eta ^{\mu }\left( t\right) \eta ^{\nu }\left( t^{\prime
}\right) \right\rangle  &=&N^{\mu \nu }\delta \left( t-t^{\prime }\right) 
\end{eqnarray*}%
where $N^{\mu \nu }$ is real symmetric and positive.

Let the eigenvalues and right/left eigenvectors of $F$ be given by 
\begin{eqnarray*}
\sum_vF_\nu ^\mu u_I^\nu &=&\lambda \left( I\right) u_I^\mu ;\quad \sum_\mu
v_\mu ^IF_\nu ^\mu =\lambda \left( I\right) v_v^I \\
I &=&1,...,\mathcal{N}
\end{eqnarray*}
with 
\begin{equation*}
\mathop{\rm Re}\lambda \left( I\right) >0.
\end{equation*}
We choose eigenvectors so they form a complete bi-orthonormal set, i.e., 
\begin{equation*}
\sum_Iu_I^\nu v_\mu ^I=\delta _\mu ^\nu \quad \quad \sum_\mu v_\mu
^Iu_J^\mu =\delta _J^I\,\,.
\end{equation*}

The Fokker-Planck equation is 
\begin{equation}
\partial _tP\left( \vec{\xi},t\right) =\sum_{\mu ,v}\left\{ \frac 12\partial
_\mu N^{\mu \nu }\partial _\nu +\partial _\mu F_\nu ^\mu \xi ^\nu \right\}
P\left( \vec{\xi},t\right) \,\,,  \label{A-FP}
\end{equation}
where 
\begin{equation*}
\partial _\mu \equiv \frac \partial {\partial \xi _\mu }
\end{equation*}
differentiates all factors to its right.

The stationary distribution is also a Gaussian 
\begin{equation}
P^{*}\left( \vec{\xi}\right) =\left( 2\pi \right) ^{-\mathcal{N}/2}\left(
\det \mathbb{G}\right) ^{-1/2}\exp \left[ -\frac 12\sum_{\mu ,v}\xi ^\mu
G_{\mu \nu }\xi ^\nu \right] \,\,,  \label{eq:PG1}
\end{equation}
where 
\begin{equation*}
\mathbb{G}=\mathbb{C}^{-1}
\end{equation*}
with $\mathbb{C}$ being the correlation matrix: 
\begin{equation*}
\left\langle \xi ^\mu \xi ^\nu \right\rangle =C^{\mu \nu }.
\end{equation*}
The explicit form of $\mathbb{C}$ is found in the standard way, except that
matrices are involved. Thus, it is formally  
\begin{equation*}
\mathbb{C}=\int \frac{d\omega }{2\pi }
\left( \frac 1{i\omega +\mathbb{F}} \right)
\mathbb{N}
\left( \frac 1{i\omega +\mathbb{F}} \right) ^{\dagger }
\end{equation*}
In the frame where $\mathbb{F}$ is diagonal, this integral is trivial. The
result is, in terms of the explicit matrix elements: 
\begin{equation*}
C^{\mu \nu }=\sum_{I,J}\frac{u_I^\mu N^{IJ}u_J^\nu }{\lambda \left( I\right)
+\lambda \left( J\right) }\,\,,  
\end{equation*}
where 
\begin{equation*}
N^{IJ}\equiv \sum_{\mu ,v}v_\mu ^IN^{\mu \nu }v_\nu ^J  
\end{equation*}
is just $\mathbb{N}$ in the new frame.

The probability currents are displayed in Equation (\ref{A-FP}):

\begin{equation*}
K^{\mu }=-\sum_{v}\left\{ \frac{1}{2}N^{\mu \nu }\partial _{\nu }+F_{\nu
}^{\mu }\xi ^{\nu }\right\} P\,\,,
\end{equation*}%
so that the \emph{stationary} currents are 
\begin{equation*}
K^{\ast \mu }\left( \vec{\xi}\right) =-\sum_{v}\left\{ \frac{1}{2}N^{\mu \nu
}\partial _{\nu }+F_{\nu }^{\mu }\xi ^{\nu }\right\} P^{\ast }\left( \vec{\xi%
}\right) \,\,.
\end{equation*}%
To see that this is the (generalized) curl of a vector field, we exploit the
Gaussian property of $P^{\ast }$ and substitute 
\begin{equation*}
\xi ^{\gamma }P^{\ast }=-\sum_{v}C^{\gamma \nu }\partial _{\nu }P^{\ast }
\end{equation*}%
into the second term. Thus, 
\begin{equation*}
K^{\ast \mu }\left( \vec{\xi}\right) =\sum_{v}\left\{ \sum_{\gamma
}F_{\gamma }^{\mu }C^{\gamma \nu }-\frac{1}{2}N^{\mu \nu }\right\} \partial
_{\nu }P^{\ast }\left( \vec{\xi}\right) .
\end{equation*}%
Next, we want to show that the matrix in $\left\{ ...\right\} $ is indeed
antisymmetric. Recall that $u$ is an eigenvector of $F$, so that 
\begin{equation*}
\sum_{\gamma }F_{\gamma }^{\mu }C^{\gamma \nu }=\sum_{I,J}\frac{\lambda
\left( I\right) }{\lambda \left( I\right) +\lambda \left( J\right) }%
u_{I}^{\mu }N^{IJ}u_{J}^{\nu }\,\,.
\end{equation*}%
For $N^{\mu \nu }$, we cast it in terms of $N^{IJ}$, i.e., 
\begin{equation*}
N^{\mu \nu }=\sum_{I,J}u_{I}^{\mu }N^{IJ}u_{J}^{\nu }\,\, ,
\end{equation*}%
so that the terms in $\left\{ ...\right\} $ combine easily. The result is
\begin{equation*}
K^{\ast \mu }=\sum_{v}A^{\mu \nu }\partial _{\nu }P^{\ast }
\end{equation*}%
with 
\begin{equation*}
A^{\mu \nu }=\frac{1}{2}\sum_{I,J}\frac{\lambda \left( I\right) -\lambda
\left( J\right) }{\lambda \left( I\right) +\lambda \left( J\right) }%
u_{I}^{\mu }N^{IJ}u_{J}^{\nu }\,\,,
\end{equation*}%
which is manifestly antisymmetric. Note that this $A^{\mu \nu }$ should not
be confused with the $A_{i}^{j}$ in Section \ref{cons}.


\begin{thebibliography}{99}
\bibitem{Einstein} In the literature, transition
rates are mostly denoted with only subscripts, e.g., $w_{ij}$. However, the
rate from $\mathcal{C}_{j}$ to $\mathcal{C}_{i}$ sometimes appear as 
$w_{ji}$ and sometimes as $w_{ij}$. We believe there is less confusion 
by using super- and subscripts, along with the mnemonic 
``transitions are like gravity, going from
top to bottom.'' This has the further advantage that matrix multiplication
now takes the usual form, e.g., $\sum_{j}W_{i}^{j}P_{j}$. We caution the
reader, however, that Einstein summation convention is \emph{never} used
here, since we frequently face, e.g., $W_{i}^{j}P_{j}$ with \emph{no sum} 
over $j$.
\end{thebibliography}
\end{document}